\documentclass[a4paper,fleqn,usenatbib]{mnras}

\usepackage{Times}

\usepackage[T1]{fontenc}
\usepackage{ae,aecompl}

\usepackage{graphicx}	
\usepackage{amsmath}	
\usepackage{amssymb}	
\usepackage{tablefootnote}
\usepackage{gensymb}

\usepackage{xcolor} 
\usepackage{soul} 
\definecolor{bluehl}{rgb}{0.75,0.75,1}

\usepackage{etoolbox}
\makeatletter
\patchcmd\@combinedblfloats{\box\@outputbox}{\unvbox\@outputbox}{}{%
   \errmessage{\noexpand\@combinedblfloats could not be patched}%
}%
 \makeatother

\title[VLT/MUSE observations of CR7]{The nature of CR7 revealed with MUSE: a young starburst powering extended Lyman-$\alpha$ emission at z=6.6} 

\author[J. Matthee et al.]{Jorryt Matthee$^{1}$\thanks{Zwicky Fellow -- mattheej@phys.ethz.ch}, Gabriele Pezzulli$^1$, Ruari Mackenzie$^1$, Sebastiano Cantalupo$^1$,  
\newauthor Haruka Kusakabe$^2$, Floriane Leclercq$^{2}$, David Sobral$^3$, Johan Richard$^4$, Lutz Wisotzki$^5$, 
\newauthor Simon Lilly$^1$, Leindert Boogaard$^6$, Raffaella Marino$^1$, Michael Maseda$^6$,
\newauthor Themiya Nanayakkara$^{6,7}$ \\
$^{1}$ Department of Physics, ETH Z\"urich, Wolfgang-Pauli-Strasse 27, 8093 Z\"urich, Switzerland\\
$^{2}$ Observatoire de Gen\`eve, Université de Gen\`eve, 51 chemin de P\'egase, 1290 Versoix, Switzerland\\
$^{3}$ Department of Physics, Lancaster University, Lancaster, LA1 4YB, UK\\
$^{4}$ Univ. Lyon, Univ. Lyon1, Ens de Lyon, CNRS, Centre de Recherche Astrophysique de Lyon UMR5574, F-69230, Saint-Genis-Laval, France\\
$^{5}$ Leibniz-Institut f\"{u}r Astrophysik Potsdam (AIP), An der Sternwarte 16, 14482 Potsdam, Germany\\
$^{6}$ Leiden Observatory, Leiden University, P.O. Box 9513, 2300 RA Leiden, The Netherlands\\
$^{7}$ Centre for Astrophysics and Supercomputing, Swinburne University of Technology, Hawthorn, VIC 3122, Australia
\\
}

\pubyear{2020}

\begin{document}
\label{firstpage}
\pagerange{\pageref{firstpage}--\pageref{lastpage}}
\maketitle

\begin{abstract}
CR7 is among the most luminous Lyman-$\alpha$ emitters (LAEs) known at $z = 6.6$ and consists of at least three UV components that are surrounded by Lyman-$\alpha$ (Ly$\alpha$) emission. Previous studies have suggested that it may host an extreme ionising source. Here, we present deep integral field spectroscopy of CR7 with VLT/MUSE. We measure extended emission with a similar halo scale length as typical LAEs at $z\approx5$. CR7's Ly$\alpha$ halo is clearly elongated along the direction connecting the multiple components, likely tracing the underlying gas distribution. The Ly$\alpha$ emission originates almost exclusively from the brightest UV component, but we also identify a faint kinematically distinct Ly$\alpha$ emitting region nearby a fainter component. Combined with new near-infrared data, the MUSE data show that the rest-frame Ly$\alpha$ equivalent width (EW) is $\approx100$ {\AA}. This is a factor four higher than the EW measured in low-redshift analogues with carefully matched Ly$\alpha$ profiles (and thus arguably HI column density), but this EW can plausibly be explained by star formation. Alternative scenarios requiring AGN powering are also disfavoured by the narrower and steeper Ly$\alpha$ spectrum and much smaller IR to UV ratio compared to obscured AGN in other Ly$\alpha$ blobs. CR7's Ly$\alpha$ emission, while extremely luminous, resembles the emission in more common LAEs at lower redshifts very well and is likely powered by a young metal poor starburst.  

\end{abstract} 
\begin{keywords}
galaxies: high-redshift -- cosmology: observations -- galaxies: evolution -- cosmology: dark ages, reionisation, first stars
\end{keywords}



\section{Introduction}
Over the last years, new deep and wide-field extragalactic surveys have resulted in the discovery of relatively rare, bright galaxies at the end stages of cosmic reionisation ($z\gtrsim6$; \citealt{Ouchi2013,Bowler2014,Matthee2015,Shibuya2017,Smit2017}). These galaxies have UV luminosities that imply star formation rates about $25-50$ M$_{\odot}$ yr$^{-1}$ and number densities around $\sim10^{-6}$ cMpc$^{-3}$. Besides being able to confirm their redshifts spectroscopically, it is also possible to spatially resolve the most luminous systems with the {\it Hubble Space Telescope} ({\it HST}) and ALMA \citep[e.g.][]{Ouchi2013,Sobral2015,Bowler2016,Matthee2017ALMA,Matthee2017SPEC,Hashimoto2018Dragons}. Additional deep spectroscopy allows the first study of the properties of the interstellar medium (ISM) and stellar populations in these galaxies \citep{Stark2015_CIV}, and enables investigations on the fraction of light that is contributed by an active galactic nucleus \citep[AGN; e.g.][]{Laporte2017b}. 

Studies based on rest-frame UV and rest-frame far infrared spectroscopy indicate that the ISM in bright galaxies at $z\gtrsim6$ is highly ionised \citep{Inoue2016,Harikane2019,Arata2020} by hard ionising sources \citep{Stark2015_CIV,Sobral2019} and contains either little dust and/or dust with a likely very high temperature \citep[e.g.][]{Faisst2018,Bakx2020}.

Moreover, luminous galaxies at $z\gtrsim6$ appear to be complex assembling systems of multiple components identified from the UV emission of their young stars \citep[e.g.][]{Ouchi2013,Sobral2015,Bowler2017,Tamura2018} and cold gas traced by far-infrared [CII]$_{158 \mu \rm m}$ line emission \citep[e.g.][]{Matthee2017ALMA,Carniani2018}. Spatially resolved studies indicate varying line-ratios and line-to-continuum ratios \citep{Carniani2017,Matthee2019ALMA,Bakx2020}. [CII] emission is also reported to be significantly more extended than the UV continuum \citep[e.g.][]{Fujimoto2019,Ginolfi2020}, possibly tracing past outflow activity \citep{Pizzati2020}. 

The Lyman-$\alpha$ (Ly$\alpha$) emission line has mostly been used to identify and confirm the redshifts of distant galaxies, but is now also starting to be used as a tool to study the gas content in and around galaxies. For example, Ly$\alpha$ halos detected around quasars and galaxies can be used to study the properties of the circumgalactic medium (CGM; e.g. \citealt{Steidel2011,Matsuda2012,Borisova2016,Wisotzki2018}), the ISM and continuum-undetected galaxy populations \citep[e.g.][]{Zheng2011,MasRibas2017}. \cite{Leclercq2017} report no correlations between the halo scale lengths and any observed galaxy properties nor redshift at $z\approx3-5$. However, \cite{Momose2014} use a stacking analysis to show that Ly$\alpha$ halos have a larger scale length at $z=6.6$ compared to $z<6$, possibly an effect of incomplete reionisation. 

Additionally, the observed spectral profile of the Ly$\alpha$ line has emerged as a promising tracer of gas kinematics and HI column density in the ISM and the related escape fraction of ionising photons \citep[e.g.][]{Verhamme2015,Izotov2018,Matthee2018}. We know in some cases (from UV continuum or [CII]; e.g. \citealt{Sobral2015,Carniani2018,Hashimoto2018Dragons}) that there are multiple components within luminous systems each with slightly distinct systemic redshifts. This makes the physical interpretation of a spatially unresolved Ly$\alpha$ spectrum difficult, making IFU observations in the rest-frame UV necessary \citep[e.g.][]{Matthee2019MUSE}. An additional advantage of integral field spectroscopy is the possibility to define a pseudo-narrowband image which width can be optimised to maximise the signal-to-noise for a given target, which facilitates the detection of emission at low surface brightness.

One of the best sources to obtain detailed resolved Ly$\alpha$ observations is the Ly$\alpha$ emitter (LAE) COSMOS Redshift 7 (CR7, $z_{\rm Ly\alpha}=6.606$; \citealt{Matthee2015,Sobral2015}), which is one of the most luminous LAEs known at $z>6$. CR7 stands out with respect to other galaxies known at this epoch because of its high Ly$\alpha$ luminosity and the tentative detection of the high ionisation HeII emission line \citep{Sobral2015,Sobral2019}, which could point towards an extremely hot stellar population and/or an AGN \citep[e.g.][]{Sobral2015,Pallottini2015,Bowler2016,Pacucci2017}. Earlier studies revealed that CR7 is a multiple component system, consisting of (at least) three UV emitting components \citep{Sobral2015} and four [CII] components (of which some overlap with UV components; \citealt{Matthee2017ALMA}). Besides [CII], metal emission through the [OIII]$_{5008}$ line is plausibly present \citep{Matthee2015,Bowler2016}, although the large point spread function (PSF) of the {\it Spitzer}/IRAC data challenges measurements of its spatial variations over the multiple components, particularly as this is degenerate with the stellar mass distribution \citep[e.g.][]{Agarwal2016}.

In this paper, we present resolved Ly$\alpha$ data from the Multi Unit Spectroscopic Explorer (MUSE; \citealt{Bacon2010}) of CR7. We investigate the origin of the Ly$\alpha$ emission in CR7, how the Ly$\alpha$ surface brightness and line profiles compare to other galaxies. We also investigate which UV and [CII] components are responsible for the Ly$\alpha$ emission and take advantage of the 3D nature of IFU data to identify kinematically distinct components within the extended Ly$\alpha$ halo. This study is allowed by the availability of new deep, ground-layer adaptive-optics (GLAO) assisted observations with the MUSE integral field unit on the Very Large Telescope (VLT). These data are analysed in conjunction with a new analysis of {\it HST} and ground-based near-infrared data with significantly improved sensitivity compared to previous works \citep[e.g.][]{Bowler2016,Sobral2019}.

The structure of the paper is as follows. In \S $\ref{sec:summary_previous}$ we first summarise earlier results and measurements on CR7 that are most relevant for our analysis. Then, in \S $\ref{sec:data}$ we describe the data used in this paper, including VLT/MUSE observations, their reduction and the reduction of archival {\it HST} data. \S $\ref{sec:uvmorph}$ presents the UV morphology. We explore CR7's Ly$\alpha$ emission in 3D in \S $\ref{sec:lya_origin}$, including the Ly$\alpha$ surface brightness profile, spatial offset compared to the UV and identification of variations in the Ly$\alpha$ line profile in the MUSE data. Spectroscopic and photometric flux measurements and the measurement of UV luminosity, slope and Ly$\alpha$ equivalent width are presented in \S $\ref{sec:fluxes}$. We discuss the spatial origin of the Ly$\alpha$ emission in \S $\ref{sec:results}$ and in \S $\ref{sec:halo_comparison}$ we compare the Ly$\alpha$ surface brightness profile of CR7 to other galaxies. Finally, we discuss the powering origin of the Ly$\alpha$ emission in \S $\ref{sec:discussion}$), focusing on comparisons to low-redshift analogues of high-redshift LAEs and on comparisons between CR7 and other bright sources of extended Ly$\alpha$ emission. Throughout the paper we use a flat $\Lambda$CDM cosmology with $\Omega_M = 0.3$, $\Omega_{\Lambda} = 0.7$ and H$_0 = 70$ km s$^{-1}$ Mpc$^{-1}$. Magnitudes are listed in the AB system \citep{OkeGunn1983}.

 \section{Earlier results on CR7} \label{sec:summary_previous}
Earlier work identified CR7 as a bright, extended LAE through Ly$\alpha$ narrow-band imaging taken with Suprime-Cam on Subaru \citep{Matthee2015}. {\it HST}/WFC3 imaging revealed three UV continuum emitting components (named A, B and C; see Fig. $\ref{fig:HSTmorph}$), of which the brightest component (A) roughly coincides with the peak Ly$\alpha$ emission \cite{Sobral2015}. Slit spectroscopy revealed a narrow Ly$\alpha$ line at $z=6.60$, which combined with the narrow-band and $Y$ band photometry resulted in a Ly$\alpha$ luminosity of $8.5\times10^{43}$ erg s$^{-1}$ and a rest-frame EW$=211\pm20$ {\AA} \citep{Sobral2015}. The luminosity and EW implied extreme ionising sources, particularly as a significant fraction of the Ly$\alpha$ emission may have been absorbed by the opaque inter galactic medium (IGM), or not seen due to the surface brightness limits of the narrow-band data. These would both indicate even higher Ly$\alpha$ luminosity and EW. 

While sensitive ALMA observations do not detect any continuum emission (indicating a low dust content), [CII]$_{158 \mu m}$ line emission is detected at various positions \citep{Matthee2017ALMA}, see the red contours in the left panel of Fig. $\ref{fig:HSTmorph}$. The brightest [CII] component overlaps with UV component A with $z_{\rm [CII]} = 6.601$. There are two nearby compact [CII] emitting sources at the position of component B with $z=6.600$ and $z=6.593$, respectively. There is no compact source of [CII] emission at the location of UV component C, but more diffuse emission is seen with a redshift $z=6.598$. For the purpose of this paper we will use the redshift of the brightest component ($z=6.601$) as the systemic redshift and as the rest-frame velocity.

The most recent analysis of the rest-frame UV spectroscopy with VLT/X-SHOOTER and the grism on {\it HST}/WFC3 has been presented in \cite{Sobral2019}, who report a $\approx3\sigma$ detection of HeII emission. The line peaks at $z=6.604$ and is located 0.8$''$ away from clump A, roughly between clumps B and C. No other rest-frame UV lines are detected. Photo-ionisation modelling indicates that the spectrum can be explained by a relatively young metal-poor starburst and does not require PopIII stars or an AGN.

\section{Data} \label{sec:data}
\subsection{VLT/MUSE}
CR7 was observed in clear conditions for 4 hours with VLT/MUSE on March 5 and 7 and April 10, 2019 as part of GTO programs 0102.A-0448 and 0103.A-0272 (PIs Cantalupo/Lilly). Each of the four observing blocks consisted of four GLAO-assisted integrations with 900s exposure times. Individual exposures were dithered randomly by $\approx2''$ and the position angle of the pointing was rotated by 90 degrees after each exposure to reduce the effects of systematics on the final datacube. 

Standard reduction steps (bias, flat-fielding, illumination correction, geometrical calibration and barycentric wavelength and flux calibration), were performed with the standard MUSE pipeline version 2.6 \citep{Weilbacher2014} implemented in {\sc ESOrex}. Additionally, we registered the astrometric frame to the GAIA DR2 reference frame by shifting the coordinates of objects within 20$''$ from the center of the MUSE field of view (FoV) to the reference catalog (assuming no geometric distortions after the standard pipeline reduction). As no object in the GAIA DR2 catalogue \citep{GAIADR2} is detected within the MUSE FoV, we use a wedding-cake approach by matching high signal-to-noise (S/N) detections in the MUSE white-light image to detections in the UltraVISTA DR4 $K_s$ band (which has been registered to the GAIA DR2 reference frame). Once the astrometry of all individual reduced cubes is matched, we use two iterations of CubEX (Cantalupo in prep.; see \citealt{Cantalupo2019} for a description) for removal of sky-line residuals, additional flat-fielding and combination of individual exposures. The white-light image of the combined cube of the first iteration was used as source-mask for the second iteration. CR7 was added manually to this mask. 

We measure the PSF at $\lambda_{\rm obs} = 925$ nm, the wavelength of CR7's Ly$\alpha$ emission, by fitting a Moffat profile to a bright star ($I=17.9$) in the field of view (FoV). The profile is best characterised with a power index $\beta=2.2$ and a full width half maximum FWHM=0.47$''$. Compared to non-GLAO observations \citep[e.g.][]{Bacon2017,Matthee2019MUSE} we find that while the core of the PSF is very narrow, the wing is somewhat more extended ($\beta=2.8$ in those non-GLAO data).

We measure the depth of the data by placing PSF FWHM-sized apertures in 67 empty sky positions identified by eye from deep {\it HST} data (see below) and the MUSE white-light image and measuring the standard deviation in the aperture-fluxes. The combined data-cube has a limiting  $5\sigma$ point-source sensitivity of $6\times10^{-19}$ erg s$^{-1}$ cm$^{-2}$ {\AA}$^{-1}$ at $\lambda_{\rm obs} = 925$ nm (including a factor 1.2 correction to account for cross-talk from the spectral resampling; see \citealt{Weilbacher2020}). We note that no strong skylines are present around CR7's Ly$\alpha$ wavelength.

   \begin{figure}
\hspace{-0.3cm} \includegraphics[width=8.7cm]{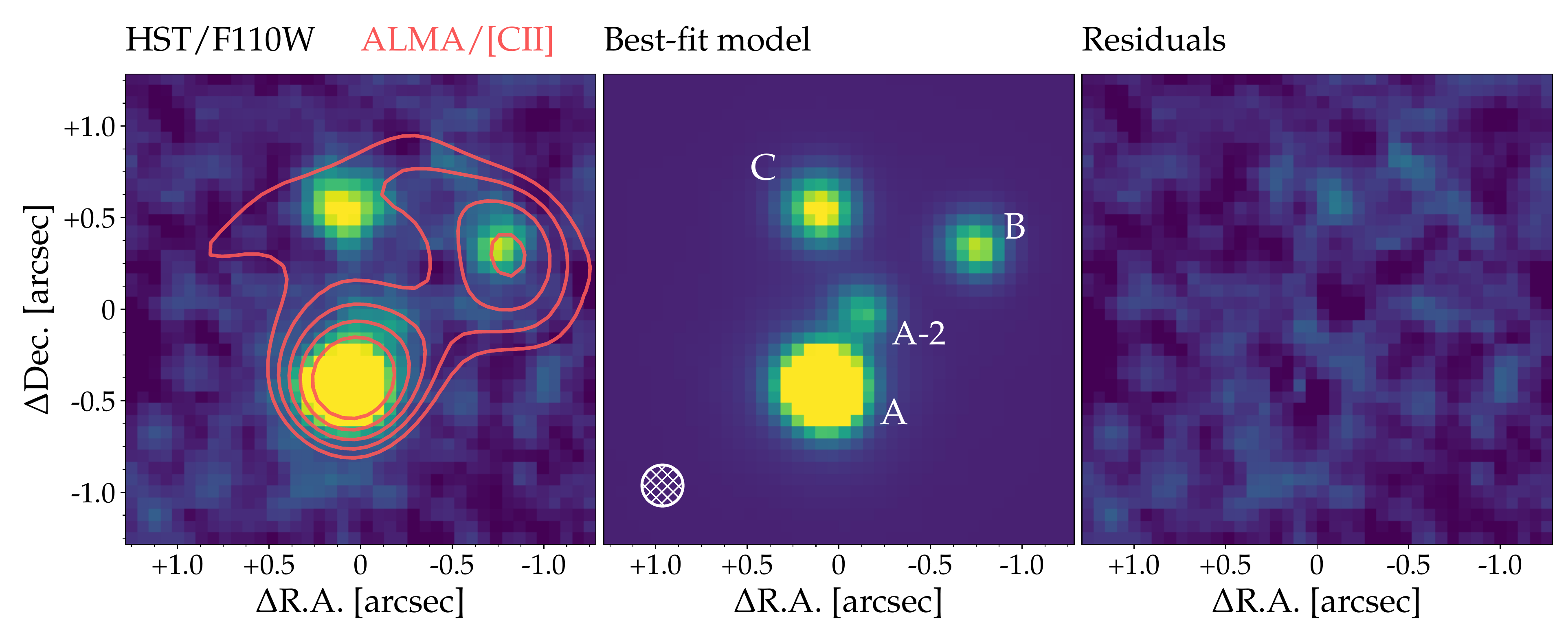}\\
\caption{Zoomed-in image of CR7's rest-frame UV emission as observed by {\it HST}/WFC3 in the F110W filter. The left panel shows the data. For illustration, light-red contours show the location of [CII] line emission as observed with ALMA. The central panel shows the best-fit model and the right panel shows the residuals after subtracting the best-fit model from the data. There are weak residuals in the centre of the main component A, which could point to a slightly steeper profile than the exponential profile used in our modelling. The PSF-FWHM of the data is shown as a white filled circle in the central panel.  }
\label{fig:HSTmorph}
\end{figure}

\subsection{HST/WFC3}
We compile all available near-infrared data in the {\sc STScI} database on CR7 observed with WFC3 on {\it HST}. The data include observations in the F110W, F140W and F160W filters. The F110W data contains 1 orbit observed in March 2012 from program 12578 (PI: Forster Schreiber) and 2 orbits observed in March and November 2017 from program 14596 (PI: Fan). The F140W contains $\approx1$ orbit worth of exposure time from grism pre-imaging in January and March 2017 from program 14495 (PI: Sobral). The F160W data consist of a total of 4 orbits obtained through the same programs as the F110W data.

The data are reduced following the method outlined in \cite{Matthee2019ALMA}. This means that individual calibrated and flat-fielded images are registered to the astrometric reference frame of the GAIA DR2 data by matching the {\it HST} detections with $K_s$ band data (see above) and finding the best astrometric solution with {\sc scamp} \citep{Bertin2006}. Finally, we use {\sc Swarp} \citep{Bertin2010} to combine individual images to a co-add with 0.064$''$ pixel scale using bilinear interpolation. Comparing the positions of objects within the central 20$''$ of the MUSE data-cube to their positions in the {\it HST/WFC3} data, we find no systematic astrometry offsets and an uncertainty of 0.02$''$ in the relative astrometry.

Using a similar method as described for the MUSE data, we measure that the {\it HST} data have 5$\sigma$ point-source sensitivities F110W = 28.2, F140W=27.3 and F160W= 27.9 and PSF FWHM $\approx0.25''$. We note that the bilinear interpolation introduces some smoothing, resulting in higher S/N at the cost of a slightly larger FWHM than the native FWHM. 
 
\subsection{Ground-based data}
We also use the most recent release (DR4) of ground-based NIR data in the $Y$, $J$, $H$ and $K_s$ bands from UltraVISTA \citep{McCracken2012} and NIR data in the $Y_{\rm HSC}$ band from the Hyper Suprime-Cam Subaru Strategic Program DR2 \citep{HSC_SURVEYPAPER}. Compared to earlier ground-based data on CR7, the most significant improvement is in the $K_s$ band data, where individual components of CR7 are now detected. The PSF FWHM of the ground-based data is $\approx0.8''$ and we measure $5\sigma$ point-source sensitivities of 26.2, 25.8, 25.5, 25.2 and 25.4 magnitudes in the $Y_{\rm HSC}$, $Y$, $J$, $H$ and $K_s$ filters, respectively.

 \section{UV morphology} \label{sec:uvmorph}
Here we aim to obtain a model that describes the UV continuum as observed with {\it HST}/WFC3 in order to have a baseline to interpret the Ly$\alpha$ morphology. We use the data in the F110W filter as these data have the best sensitivity.\footnote{The F110W filter contains the Ly$\alpha$ emission line; this contribution is however weak (0.03 magnitude in an aperture integrated over component A). We have checked that the morphology as measured in the F160W filter is fully consistent with the results obtained from the F110W filter within the 1$\sigma$ uncertainties.}

Following earlier work on the morphologies and sizes of high-redshift galaxies \citep[e.g.][]{Shibuya2015,Bowler2017,PaulinoAfonso2018}, we use exponential profiles (i.e. Sersic profiles with $n=1$). For simplicity and to limit the number of free parameters, we assume circularly symmetric light-profiles. We use the following general parametrisation:
\begin{equation}
I(a) =  I_{\rm eff} \exp({-b_n [(\frac{a}{r_{\rm eff}})^{1/n} -1]}),
\end{equation}
with $n$ the Sersic index (set to $n=1$ for an exponential profile), and $b_n$ is calculated from the incomplete gamma function (see \citealt{Erwin2015}) such that $r_{\rm eff}$ is the effective (half-light) radius and $I_{\rm eff}$ is the surface brightness at the effective radius. We note that for an exponential profile the half-light radius is related to the scale length as r$_{\rm eff}\approx1.67835 \, r_s$ where $I(a)\propto \exp{(-a/r_s)}$.

Interestingly, besides the 3 previously known components, our new deeper {\it HST} data reveal weak clumpy emission somewhat north of clump A of CR7 (Fig. $\ref{fig:HSTmorph}$). This additional flux (named A-2) is seen in both F110W and F160W data, indicating it is continuum emission. The integrated S/N of component A-2 is 3.8. Here we model A-2 as an additional point-source for simplicity. We note that if we would allow the Sersic index or the ellipticity to vary in clump A instead of adding an extra component would not result in a good fit.

 \begin{figure*}
\hspace{-0.4cm}\includegraphics[width=17cm]{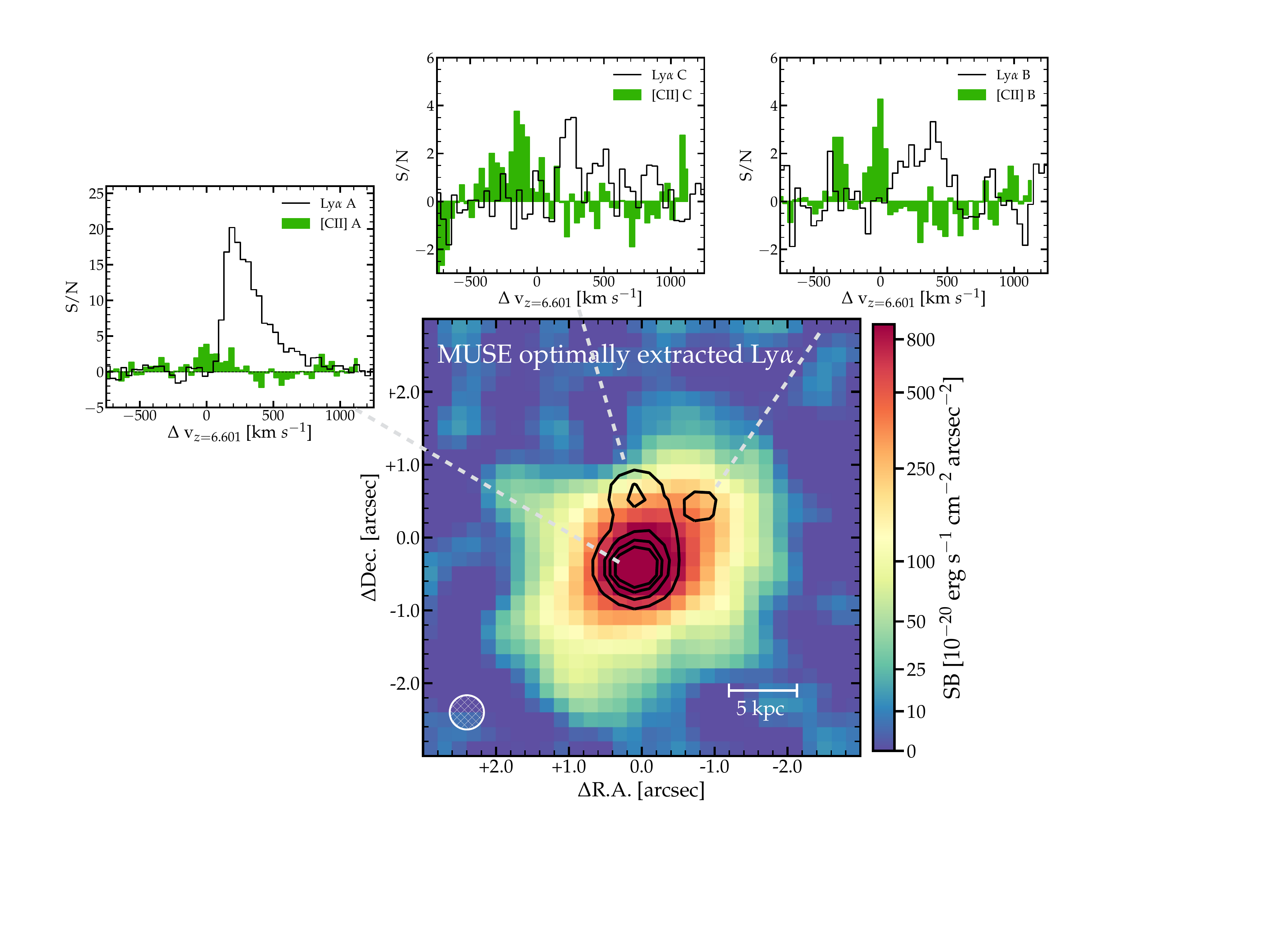}\\  
\caption{Overview of the MUSE Ly$\alpha$ data on CR7. The large panel shows an optimally extracted Ly$\alpha$ image with logarithmic colour scaling to emphasise both extended emission and the location of the peak emission. The image is smoothed with a gaussian kernel with $\sigma=0.2''$. The black contours show the {\it HST}-based UV continuum image (convolved to match the PSF of the MUSE data, which is shown as white circle in the bottom-left of the panel). The three outset panels show the Ly$\alpha$ profile (black lines; see \S $\ref{sec:lineprofile}$) extracted in PSF-sized apertures at the locations of the three UV components. We also show the [CII] spectra at the same locations as observed by ALMA (\citealt{Matthee2017ALMA}) in green. The spectra are shifted to the rest-frame velocity of the [CII] emission in component A.}
\label{fig:1}
\end{figure*}

We model CR7's UV continuum emission using a combination of 2 exponential profiles (clumps A and C) and 2 point sources (clumps A-2 and B, which are unresolved). We fit this morphological model with 14 free parameters (8 for the centroids of the four components, 2 for the total fluxes of clumps A-2 and B and 4 for the effective radii and the normalisations of clumps A and C) using {\sc imfit-mcmc} \citep{Erwin2015}, which simultaneously accounts for PSF convolution and pixel-based noise properties based on the propagated {\it HST} weight image. We re-normalised the weight image to certify that the noise measured in PSF-sized apertures in the noise map is in agreement with the value measured using empty aperture measurements on the real data. {\sc imfit-mcmc} uses a differential evolution implementation of Monte Carlo Markov Chain (MCMC) \citep{Vrugt2009} and the same number of Markov chains as the number of free parameters. We use 5000 iterations in the burn-in phase. Chains are run for a maximum of 100,000 generations, although we note convergence is typically reached after $\approx30,000$ iterations. 

Initial parameter guesses were obtained from running a single iteration of {\sc imfit} that uses the Levenberg-Marquardt algorithm to find the best-fit parameters using a Poisson Maximum Likelihood statistic (see \citealt{Erwin2015} for details and comparisons to $\chi^2$). Flat priors with wide boundary conditions are applied. The only boundary condition that is important applies to the central position of the four components, which are allowed to vary by 1 pixel (i.e. $3\times$ the $1\sigma$ astrometric uncertainty). Within these boundary conditions, the results are well converged to a single local maximum in the likelihood space. Then, we use the median and 16th-84th percentiles of the marginalised posterior distribution to find the best-fit parameters and their uncertainties. We measure effective radii r$_{\rm eff} = 0.30^{+0.12}_{-0.07}$ kpc and r$_{\rm eff} = 0.36^{+0.36}_{-0.17}$ kpc for clumps A and C, respectively in the F110W data, but note that care must be taken in interpreting the size of clump A due to the nearby clump A-2. The distance between the center of A and A-2 is $2.2\pm0.4$ kpc ($\approx0.41''$). For the F110W data, the best-fit model and the residual image are shown in Fig. $\ref{fig:HSTmorph}$. The measurements for F160W are consistent within the 1$\sigma$ uncertainties. We note that the contribution from clump A-2 to the total A+A-2 flux is comparable in the F110W and F160W filters ($\approx10$ \%).

 \section{CR7's Ly$\alpha$ emission in 3D} \label{sec:lya_origin}
In this section we use the advantage of the 3D data to optimally measure the morphology of the Ly$\alpha$ emission (\S $\ref{sec:lyamorph}$) and spatial offsets to the UV continuum (\S $\ref{sec:astrometry})$. We also explore spatial variations in the spectral line profile (\S $\ref{sec:lineprofile}$)  and use those to unveil a second faint source of Ly$\alpha$ emission within the system (\S $\ref{sec:2ndline}$).

In Fig. $\ref{fig:1}$ we show, for illustrative purposes, the Ly$\alpha$ image of CR7. We also show extracted 1D Ly$\alpha$ (MUSE) and [CII] (ALMA; \citealt{Matthee2017ALMA}) spectra in various locations extracted in PSF-sized apertures. These will be discussed in \S $\ref{sec:lineprofile}$. The Ly$\alpha$ image is an optimally extracted image obtained from the collapse of a three dimensional segmentation mask (see \citealt{Borisova2016} for details). As the number of wavelength-layers and hence the noise properties vary per pixel we do not use this image for quantitative measurements. For a proper comparison, we show contours of the UV continuum based on the best-fit intrinsic UV morphology convolved with the PSF of the MUSE data (see \S $\ref{sec:uvmorph}$). Fig. $\ref{fig:1}$ shows that CR7's Ly$\alpha$ emission is rather smooth and peaks close to the main UV continuum emitting component (clump A), while it extends over $\approx4''$ in diameter, covering the other UV components (clumps B and C) in agreement with narrow-band data \citep{Sobral2015}. Ly$\alpha$ emission appears elongated in the direction of clump B, the component that is faintest in the UV continuum.

 \begin{figure*}
\includegraphics[width=16.3cm]{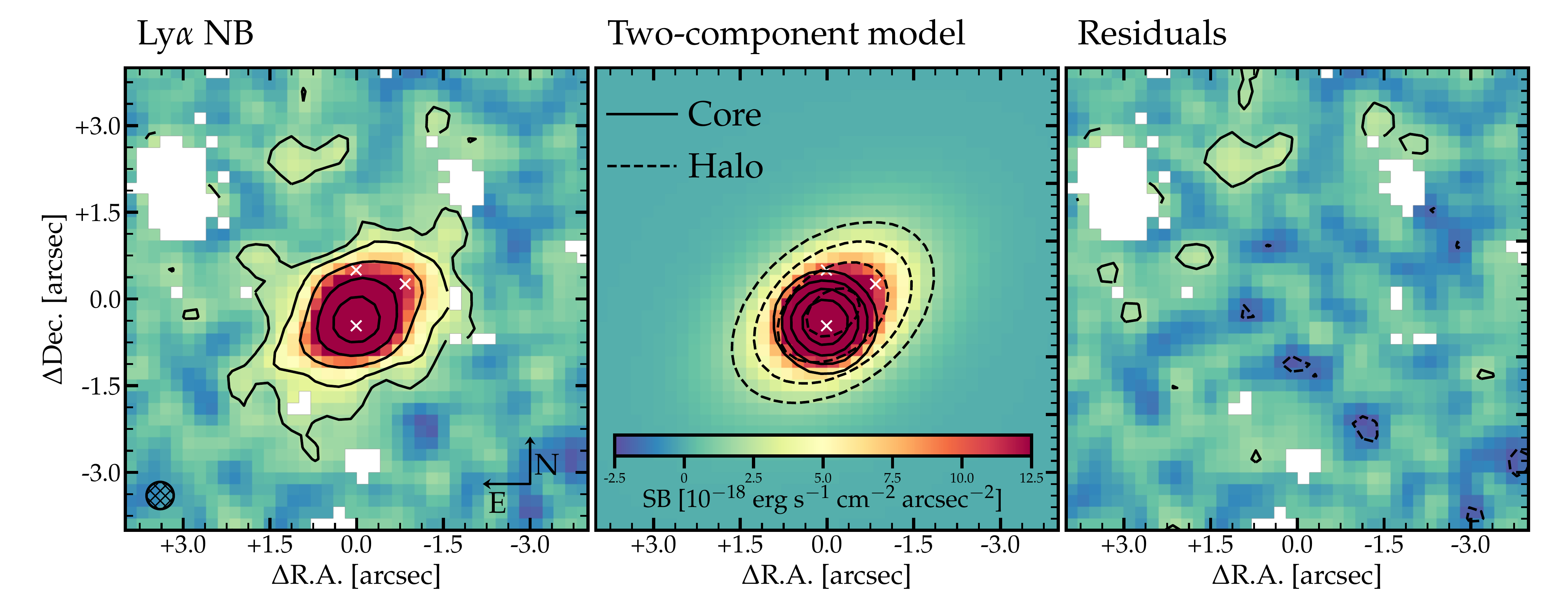} 
\caption{Zoomed-in images of CR7's continuum-subtracted Ly$\alpha$ emission as constructed with a pseudo-narrowband from the MUSE data. The left image shows the data, where the contours correspond to the $2, 4, 8, 16\sigma$ levels. Pixels with continuum emission in the {\it HST} data (besides CR7 itself) are aggressively masked and shown in white. The PSF-FWHM is illustrated as a black hashed circle in the bottom left. The middle panel shows the best-fit two-component model with an exponential halo. The contours are drawn at the same levels as in the left panel and are drawn for both the core component (solid lines) and halo component (dashed lines). The right panel shows that the best-fit model results in no substantial residuals.}
\label{fig:MUSE_BEST_MCMC}
\end{figure*}

\subsection{Ly$\alpha$ morphology}  \label{sec:lyamorph} 
Here, we focus on describing the morphology of CR7's Ly$\alpha$ emission, following the same method applied to the rest-frame UV imaging (i.e. using {\sc imfit-mcmc}, see \S $\ref{sec:uvmorph}$). We create a Ly$\alpha$ pseudo-narrowband  image by collapsing over 12 layers from $\lambda_{\rm obs}=9242-9255$ {\AA} (from $\approx -100$ to $+350$ km s$^{-1}$ with respect to the peak of the Ly$\alpha$ emission). The continuum is subtracted using a pseudo-narrowband with same width from $\lambda_{\rm obs}=9284-9297$ {\AA}  ($\approx$ +1300 to +1750 km s$^{-1}$ with respect to the Ly$\alpha$ peak), but we note this has a negligible effect due to the high observed EW. We also create a noise image based on propagating the variance cube provided by the MUSE pipeline. The noise image is re-normalised to the level measured from empty-sky pixels in the narrow-band image. We aggressively mask pixels where there is a continuum detection of a foreground source in the {\it HST} data, as shown in the left panel of Fig. $\ref{fig:MUSE_BEST_MCMC}$.

As shown by the UV continuum contours in Fig. $\ref{fig:1}$ (which are convolved to have the same PSF-FWHM as the MUSE data with FWHM=0.47$''$), it is clear that the Ly$\alpha$ morphology is significantly different from the UV morphology. Following the methodology from \cite{Wisotzki2015}, we describe the Ly$\alpha$ morphology as a combination of a (PSF-convolved) UV continuum model and an extended component. The UV continuum-like model is named the `core' component from now on, while we name the extended component the `halo' component. We model the halo component with an exponential profile and allow for non-circularly symmetric light distributions by also fitting for ellipticity and the position angle. Ellipticity is defined as $\epsilon = 1 - b/a$, where $a$ and $b$ are the semi-major and semi-minor axes respectively. The {\it circularised} radius is related to these axes as $r_{\rm circularised}=\sqrt{a b}$. We note that we have experimented fitting the halo with a Sersic model with $n\neq1$, but found that those fits do not converge without imposing a strong prior on $n$.

The difficulty in modelling CR7's Ly$\alpha$ emission is that there are potentially three core components as CR7 consists of (at least) three UV emitting components. Fig. $\ref{fig:1}$ however clearly shows that any Ly$\alpha$ emission from clumps B and C appears subdominant in the total Ly$\alpha$ image. We therefore do not include `core' Ly$\alpha$ emission at the positions of clumps B and C, and note that including such components would result in a worse reduced $\chi^2$.\footnote{Ly$\alpha$ emission with a distinctly different Ly$\alpha$ profile from the majority of Ly$\alpha$ emission is observed around the position of UV component B (\S $\ref{sec:2ndline}$). This component however has a negligible flux and does not impact the overall morphology. We have verified this by analysing a Ly$\alpha$ pseudo-narrowband collapsed over a narrower wavelength range that does not contain the additional redder Ly$\alpha$ component. These results are fully consistent within the uncertainties.}

Another possible complication arises from the faint clump A-2 that we have discovered close to the main clump A, see \S $\ref{sec:uvmorph}$, particularly as the MUSE data do not resolve these substructures. As clump A-2 is more than 10 times fainter than clump A we choose to model the Ly$\alpha$ core emission by using component A only, but we have verified that our results are unchanged within the uncertainties when incorporating component A-2 as well (but fixing the relative luminosities of components A and A-2 to the relative luminosity in the F110W data). 

Hence, our two-component model of the Ly$\alpha$ emission includes a circularly symmetric exponential component with $r_{\rm eff} = 0.30$ kpc centred on the position of clump A and an extended halo-component. The position of the core component is allowed to vary by 2$\sigma_{\rm astrometry}$, where $\sigma_{\rm astrometry} = 0.024''$, the uncertainty in the relative positions of objects in the MUSE and {\it HST} data (\S $\ref{sec:data}$). The normalisation of the core is a free parameter. The position, scale radius and normalisation of the halo-component are allowed to vary freely. The fitted parameters and their 68-percentile confidence intervals for the two-component model are listed in Table. $\ref{tab:lya_fits}$. The exponential halo is characterised by a scale radius of $3.0^{+0.3}_{-0.3}$ kpc and contributes more than half of the total (integrated) Ly$\alpha$ emission, see the halo flux fraction listed in Table $\ref{tab:lya_fits}$ that was derived from the posteriors. We note that forcing the core and the halo to be at the same positions (within 2 times the astrometric uncertainty) results in a best-fit with clear residuals in the centre of CR7 and worse $\chi^2$. 

\begin{table}
\caption{Best-fit parameters in our morphological core+halo model of CR7's Ly$\alpha$ emission. }
\begin{tabular}{lr}
Property & Measurement  \\ \hline
HST `A' + Exponential halo (\S $\ref{sec:lyamorph}$) & Full NB  imfit-MCMC  \\
PA & $127^{+4}_{-4}$$^{\circ}$ \\
$\epsilon$ &  $0.46^{+0.04}_{-0.04}$ \\
r$_{\rm s, halo}$ & $3.0^{+0.3}_{-0.3}$ kpc \\  
Halo flux fraction & $71^{+2}_{-2}$ \% \\ 
Distance Ly$\alpha$ - UV & $1.2^{+0.2}_{-0.2}$ kpc \\ \hline 

\end{tabular} \label{tab:lya_fits} 
\end{table}

\subsection{Positional offsets between UV and Ly$\alpha$} \label{sec:astrometry}
As described in \S $\ref{sec:lyamorph}$, we allow for positional offsets between the compact (core-like) Ly$\alpha$ emission, centred on the peak of the UV emission, and the Ly$\alpha$ halo. Here we explore whether such offset is real. 

\begin{figure}
\includegraphics[width=8.2cm]{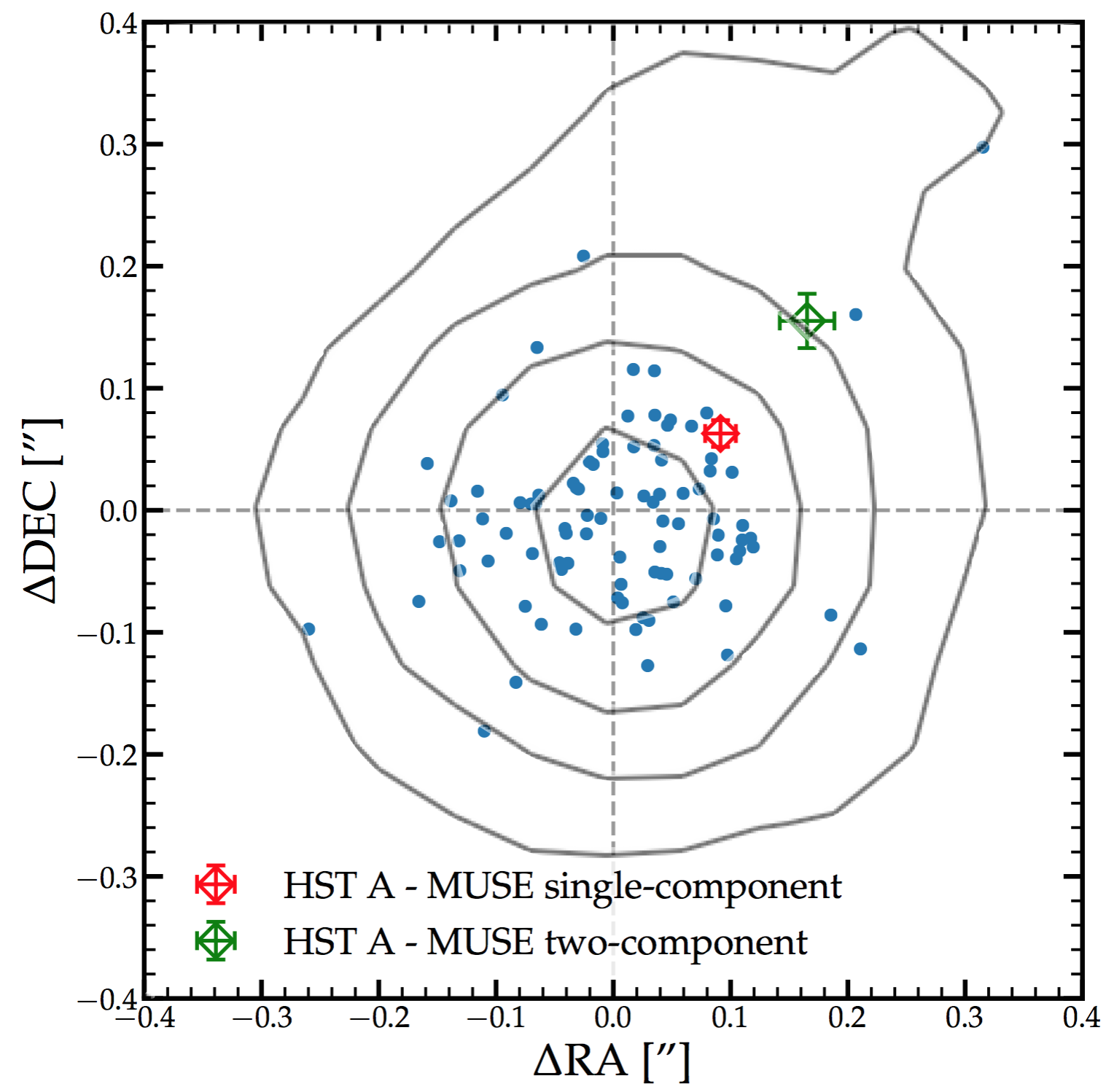}
\caption{Relative offsets between the MUSE and {\it HST}/WFC3 data. The blue points show the difference between the {\it HST} and MUSE position for all sources within 20$''$ from CR7. The 0,0 position is the centroid of UV component A. The red diamond illustrates the position of the centre of the Ly$\alpha$ emission when modelled with a single elongated component. The green diamond illustrates the position of the peak of the extended `halo'-like Ly$\alpha$ emission in the best-fit two-component model. Error-bars include the systematic uncertainty on the relative astrometry. For illustration, the contours of the F110W data on CR7's main UV component are shown in the background.} 
\label{fig:WCS}
\end{figure}

As described in \S $\ref{sec:data}$ and now shown in Fig. $\ref{fig:WCS}$, we have tested the relative astrometry between the {\it HST} and the white-light image of the MUSE data. There are no systematic offsets between the centroids of the 39 objects detected with S/N$>5$ in both images within a radius of 20$''$ from CR7. The standard deviation of the relative offsets is $0.08''$ in both the right ascension and declination directions. In Fig. $\ref{fig:WCS}$ we also show the relative positions between the centre of the UV emission and the peak of the extended Ly$\alpha$ emission when fit with a single-component exponential model (red; where we fitted a single elongated exponential light distribution similarly as described in \S $\ref{sec:lyamorph}$) and the two-component model (green). In the background, we show contour levels drawn on the {\it HST} F110W image for illustration. The relative offset between the UV and the single-component Ly$\alpha$ emission is $0.11\pm0.01''$ (modelling uncertainties), which corresponds to $\approx0.6$ kpc at $z=6.6$. The relative offset between the UV and extended Ly$\alpha$ emission is significantly larger ($0.22\pm0.4''$, corresponding to $1.2\pm0.2$ kpc) in the two-component model. Interestingly, the direction of the relative offsets of the UV and Ly$\alpha$ emission is the same as the direction towards clump A-2 and the other UV components (see \S $\ref{sec:uvmorph}$ and Fig. $\ref{fig:HSTmorph}$).

\subsection{Line profile variations} \label{sec:lineprofile} 
As illustrated in Fig. $\ref{fig:1}$, the Ly$\alpha$ line profile appears to vary throughout the system. In Appendix $\ref{sec:pseudoslit}$ we show the spatial dependence of the line-profile with pseudo-2D slit spectra extracted at various locations and with various position angles from the 3D data. Here, we explore in a model-dependent way how the Ly$\alpha$ line profile varies within the system. In this model, we parametrise the line-profile with a skewed gaussian profile \citep[e.g.][]{Shibuya2014}:
\begin{equation}
    f(v) =  {A}\,{\rm exp} \Big( -\frac{(v-v_0)^2}{2{(a_{{\rm asym}}\,(v-v_0)+d)}^2} \Big),
\end{equation} 
where $A$ is the normalisation, $v_0$ is the velocity with respect to Ly$\alpha$ peak at $z=6.601$,  $a_{\rm asym}$ the asymmetry parameter. The parameter $d$ controls the line-width and is related to the full-width half maximum as FWHM $= \frac{2 \sqrt{2 \ln 2} d}{1- (2\ln 2) {\rm a_{\rm asym}}^2}$. We convolve this line profile with the line spread function of the MUSE data, which is characterised by a gaussian profile with FWHM=70 km s$^{-1}$ at the redshifted Ly$\alpha$ wavelength \citep{Bacon2017} when we fit the line-profile to the data.

\begin{figure*}
\includegraphics[width=16.3cm]{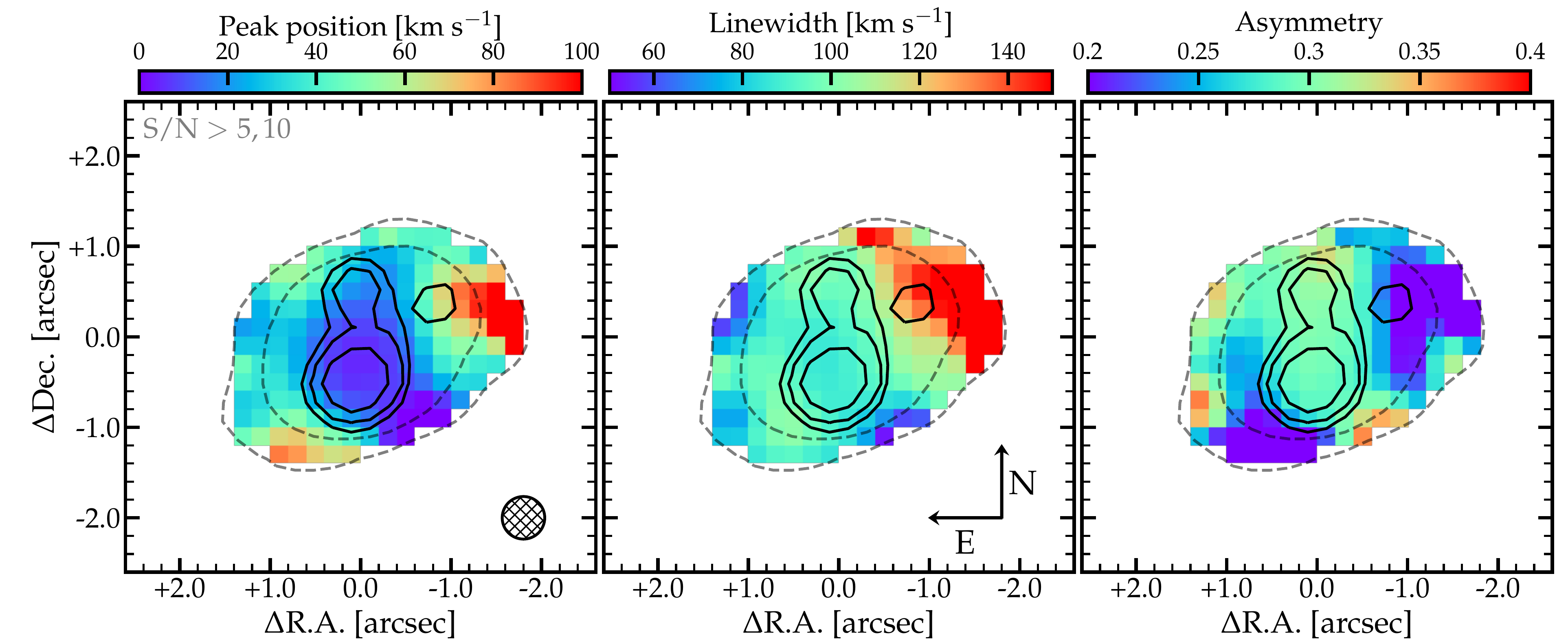} \\
\caption{Results from pixel-based fits to CR7's Ly$\alpha$ line profile. In the left panel the colour-coding corresponds to the peak position. The middle panel shows the line-width and the right panel the asymmetry. Black contours illustrate the UV morphology convolved to the MUSE PSF (illustrated as a hashed circle in the left panel). Grey dashed lines show the 5 and 10$\sigma$ contours of the continuum-subtracted collapsed narrow-band image from $-100$ to $+350$ km s$^{-1}$ with respect to the global Ly$\alpha$ peak. The Ly$\alpha$ line is best-fitted by a strongly asymmetric, relatively narrow gaussian in most locations, except for the north-west where a broader, more symmetric and redder fit is preferred. The profile in the south-eastern part is also somewhat more symmetric.}
\label{fig:MUSE_RESOLVED_LINE}
\end{figure*}

Since the interpretation of standard moment maps is not intuitive for strongly asymmetric lines, we use a pixel-based fitting approach for our spatially resolved analysis. First, we smooth the MUSE data with a gaussian with $\sigma=1.5$ pixel ($0.3''$) to improve the S/N. Then, for each pixel within the $5\sigma$ contours of the Ly$\alpha$ narrow-band image (e.g. Fig. $\ref{fig:MUSE_BEST_MCMC}$), we extract the 1D spectrum from -750 to +1500 km s$^{-1}$ with respect to $z=6.601$. We also extract 1D spectra in all empty sky pixels identified in \S $\ref{sec:data}$ and compute the standard deviation to measure the uncertainty in each wavelength-layer. Finally, we use the {\sc python} package {\sc lmfit} to find the best-fit combination of $A$, $v_0$, $a_{\rm asym}$ and $d$ for each pixel. We note that because of smoothing and because of the PSF the results between neighbouring pixels are somewhat correlated. The pixel-based results are shown in Fig. $\ref{fig:MUSE_RESOLVED_LINE}$. It can be seen that the fitted line becomes particularly redder, broader and more symmetric in the north-western part (i.e. around clump B) and similarly (but to a smaller extent and in a lower S/N region) in the southern part. Besides, in general it appears that the peak position is somewhat more redshifted in the outskirts of the system than in the centre around clump A.

\begin{figure}
\includegraphics[width=8.4cm]{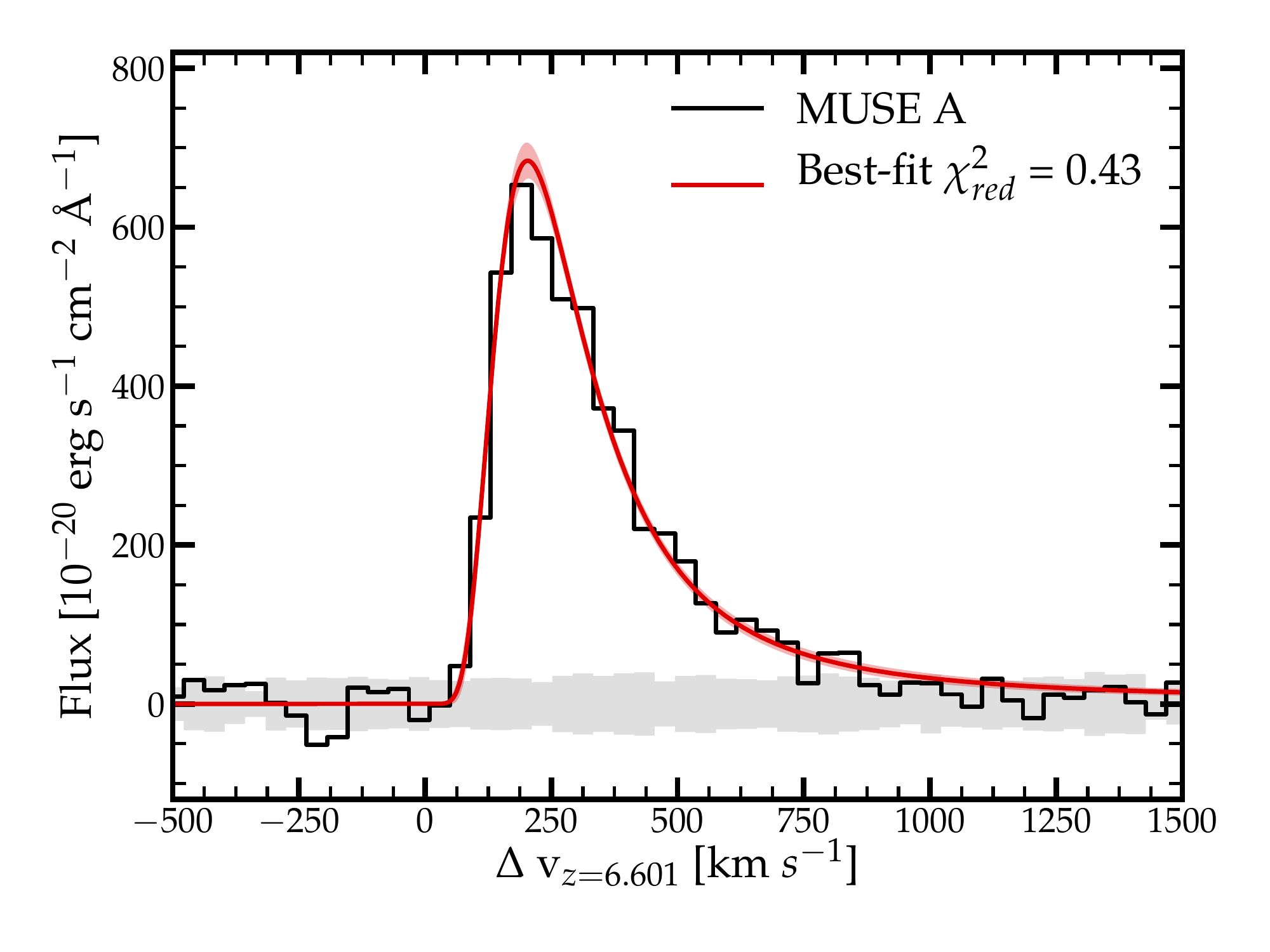}\\
\includegraphics[width=8.4cm]{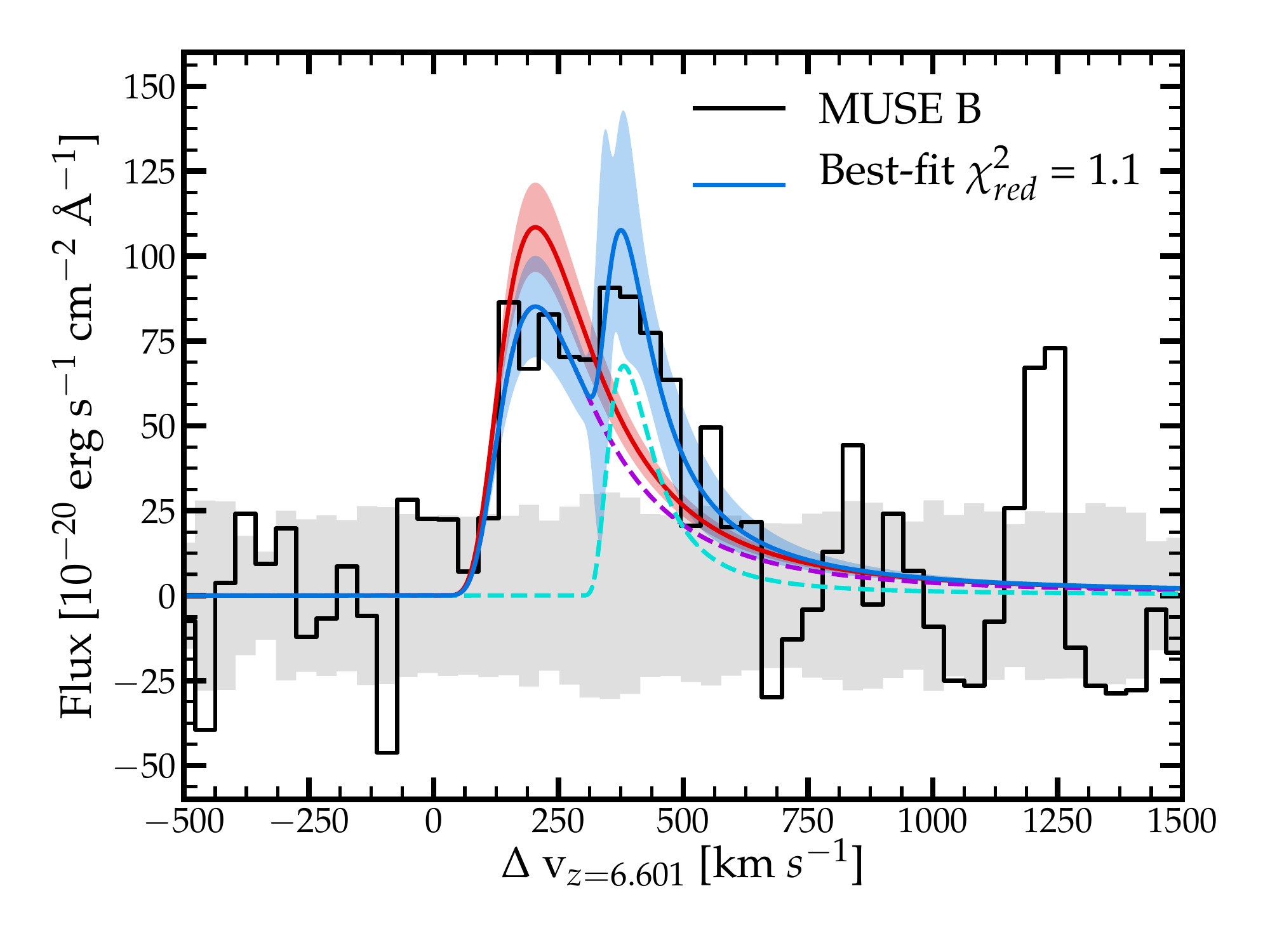}\\
\caption{Extracted one-dimensional Ly$\alpha$ spectra at the locations of the peak Ly$\alpha$ flux (top panel) and at the location of the reddest peak position (i.e. slightly west of clump B; bottom panel). The red line and shaded region show the best-fitted single skewed gaussian model and its 68\% confidence interval. The blue line and its shaded region show the best-fitted double skewed gaussian model, where the shape of the bluer component is fixed to the shape of the Ly$\alpha$ line shown in the top panel. The purple and cyan dashed lines show the individual lines that are part of the two-component fit. The grey shaded region shows the 1$\sigma$ noise level. }
\label{fig:MUSE_1D_LINES}
\end{figure}

To further explore the origin of the line-profile variations, we show two example 1D spectra (and their best-fits) in Fig. $\ref{fig:MUSE_1D_LINES}$. The top panel shows the Ly$\alpha$ line at the peak Ly$\alpha$ emission (close to the peak UV emission; see \S $\ref{sec:astrometry}$), while the bottom panel shows the Ly$\alpha$ line extracted in the region with the reddest peak position (i.e. around clump B). At peak emission, the Ly$\alpha$ line is very well described by a skewed gaussian with $v_0 = 204\pm4$ km s$^{-1}$ with respect to $z_{\rm sys}=6.601$, $a_{\rm asym}=0.285\pm0.014$ and FWHM$=246\pm12$ km s$^{-1}$. Around clump B, the Ly$\alpha$ line appears much broader without a clear single peak.

\subsection{A second Ly$\alpha$ emitting component} \label{sec:2ndline}
We have noticed that the line-shape in the region near UV component B is different from the rest of the Ly$\alpha$ halo. We therefore hypothesise that there are two Ly$\alpha$ emission lines (separated by roughly 200 km s$^{-1}$) at this position. Indeed, we find that a two-component fit is preferred over a single component with the same shape as component A ($\chi^2_{r} = 1.1$ versus $\chi^2_{r} = 1.5$). Indications for a second Ly$\alpha$ emitting component are also seen in pseudo-2D slit spectra shown in Appendix $\ref{sec:pseudoslit}$. We show in Fig. $\ref{fig:MUSE_1D_LINES}$ that it is possible to fit the profile as a combination of two skewed gaussians, where we fix the peak position, asymmetry and FWHM of the bluer component to those of the Ly$\alpha$ line at the peak flux position, we require a minimum peak separation of 100 km s$^{-1}$ and we fix the asymmetry of the redder components to the asymmetry measured at the peak flux. We find that the second peak is redshifted by $177\pm24$ km s$^{-1}$ with respect to the main Ly$\alpha$ component, and has a FWHM$=114\pm90$ km s$^{-1}$.  

\begin{figure*}
\includegraphics[width=16.6cm]{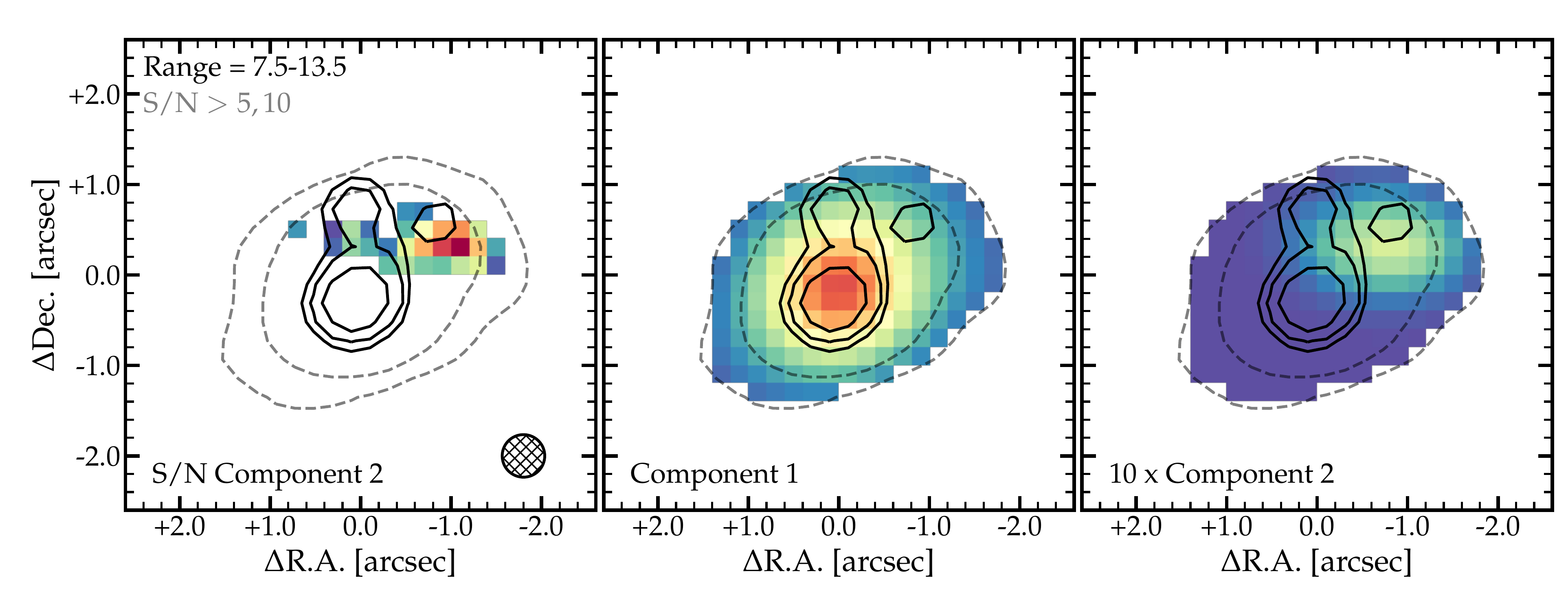}\\
\caption{The locations where the Ly$\alpha$ line is preferably fitted with a two-component skewed gaussian model. The left panel shows the S/N in the pixels in which the S/N of the second component is $>7.5$. The middle panel shows the integrated flux of the main spectral component by integrating over the velocity axis and the right panel shows the integrated flux of the second spectral component (multiplied by a factor 10 for visibility). We also illustrate the PSF of the MUSE data (hashed circle in the left panel), the rest-frame UV contours (black solid lines) and the S/N contours of the total Ly$\alpha$ narrow-band image (grey dashed lines). The central panel shows that the elongation of the main kinematic component of the Ly$\alpha$ emission is still elongated. }
\label{fig:MUSE_TWO_LINECOMPONENT}
\end{figure*}

We generalise this method to our resolved pixel-based fitting and re-fit the 1D spectrum in each pixel both with a single skewed gaussian and a combination of two skewed gaussians where we fix the shape of the bluer line to the shape of the Ly$\alpha$ line at the peak flux, pose a minimum on the separation of the two lines and fix the asymmetry of the redder component as described above. For each fit, we calculate the difference in reduced $\chi^2$ for the single and two-component fits and also measure the S/N of the red component (for example the S/N of the cyan line in the right panel in Fig. $\ref{fig:MUSE_1D_LINES}$). From visual inspection of the fits, we determine that a second component is robustly fitted when the S/N of the second line is higher than 7.5 and the reduced $\chi^2$ is improved. The left panel of Fig. $\ref{fig:MUSE_TWO_LINECOMPONENT}$ shows the pixels at which these two criteria are simultaneously met. Note that due to the additional S/N requirement, second components could in practice only be identified within the 10$\sigma$ contour levels of the total Ly$\alpha$ narrow-band image. 

As illustrated by the middle and right panels of Fig. $\ref{fig:MUSE_TWO_LINECOMPONENT}$, the integrated flux of the second component is much fainter than that of the main component, even at the location where the second component peaks. The second Ly$\alpha$ emitting component (peaking at $z=6.6105$) has a Ly$\alpha$ luminosity of only $(9\pm2) \times10^{41}$ erg s$^{-1}$, which is $\approx2$ \% of the total Ly$\alpha$ flux. The Ly$\alpha$ EW of this component is moderate EW ($\approx20$ {\AA}, see \S $\ref{sec:measure_EW}$). The spatial offset between Ly$\alpha$ component 2 and the nearby UV component B should be taken with caution as we cannot exclude that the second Ly$\alpha$ component extends further to the north, where the S/N of the Ly$\alpha$ data is relatively low. We note that the tentative HeII line-emission observed in CR7 also peaks around this spatial location \citep{Sobral2019}.

There are two [CII] emitting components at $z=6.600$ and $z=6.593$ \citep{Matthee2017ALMA} that are spatially nearby Ly$\alpha$ component 2. If we interpret one of these two redshifts as systemic, then the peak of the second Ly$\alpha$ component would correspond to a velocity shift of $+414\pm24$ km s$^{-1}$ and $+689\pm24$ km s$^{-1}$, respectively. The most likely association is the one with the smaller velocity offset as it also has a smaller spatial separation between Ly$\alpha$ and [CII]. Nonetheless, both these velocity offsets are relatively high compared to the velocity offset measured at the peak of Ly$\alpha$ which is $\Delta v_{\rm component \, A} = +204\pm4$ km s$^{-1}$ and also compared to other galaxies at $z\approx5-7$ (typically $\approx+200$ km s$^{-1}$; \citealt{Matthee2019MUSE,Cassata2020}). On the other hand, these offsets are not unseen in LAEs at $z\approx2-3$ \citep[e.g.][]{Erb2014}. Regardless, the contribution of this component to the total Ly$\alpha$ flux is minimal. 

We note that we do not find additional LAEs around CR7 in the MUSE data-cube, see Appendix $\ref{app:neighbours}$.

\section{UV luminosity and colours} \label{sec:fluxes}
\subsection{Spectrophotometry}
We use the MUSE data and the newest ground-based and {\it HST}/WFC3 data to measure the Ly$\alpha$ equivalent width, the UV luminosity and the UV slope of CR7 as a whole and for its three UV components individually. We use circular 2$''$ diameter apertures for the total photometry and $0.5''$ or $0.8''$ diameter apertures for resolved photometry of the three components in the space-based/ground-based imaging data. These aperture sizes were chosen as a compromise between optimising the S/N, minimising contamination and blending and minimising the aperture corrections. We use $0.5''$ diameter apertures for the resolved MUSE measurements that are used to correct the F110W photometry for the Ly$\alpha$ contribution. 

Aperture corrections are derived for each relevant measurement by convolving the best-fit morphological model of the {\it HST}/WFC3 F110W data (see \S $\ref{sec:uvmorph}$)\footnote{The results are unchanged when best-fit F160W model is used.} with the PSF of the data that is measured using {\sc Imfit}. The exception is the total Ly$\alpha$ flux measurement from the MUSE data, for which we base the aperture correction on the best-fit two-component model of the MUSE data (\S $\ref{sec:lyamorph}$). Typical corrections for the total magnitude are smaller than a factor $1.2$, while corrections for resolved photometry are a factor $\approx1.3-1.8$ for the {\it HST} and MUSE data and a factor $\approx2-2.5$ for the ground-based data (where larger apertures would have been more susceptible to blending and significantly lower S/N). We list the total photometry in Table $\ref{tab:totalphot}$ and resolved photometry in Table $\ref{tab:resolvedphot}$. For consistency with previous works we combine the models of A and A-2 and present their combined photometry.
 
The total Ly$\alpha$ flux corresponds to a luminosity $(5.34\pm0.11)\times10^{43}$ erg s$^{-1}$ ($\approx5\times L^{\star}$; \citealt{Matthee2015,Konno2018}). This is a factor 1.5 smaller than the Ly$\alpha$ luminosity estimated in \cite{Sobral2015}, see \S $\ref{sec:discuss_EW}$ for a discussion.
 
Comparing the photometry to earlier photometry presented in \cite{Sobral2015,Bowler2016,Sobral2019} we find broad agreement within the 15 \% level and within the 2$\sigma$ uncertainties. Differences are driven by improved sensitivity of the newer UltraVISTA and {\it HST} data used in this work and by the use of aperture-corrections based on the measured exponential profiles for clumps A and C, instead of assuming them to be point-sources.

\begin{table}
\caption{CR7's total photometry measured with 2$''$ diameter apertures including aperture corrections based on the {\it HST} morphology (broad-band filters) and MUSE morphology (Ly$\alpha$ flux).}
\centering
\begin{tabular}{lcc}
Name & $\lambda_{c, \rm obs}$ [nm] & Measurement \\ \hline
f$_{\rm Ly\alpha}$ & 924 &  $10.74^{+0.29}_{-0.29} \times10^{-17}$ erg s$^{-1}$ cm$^{-2}$ \\
F110W & 1120 & $24.52^{+0.09}_{-0.09}$ \\
F140W & 1374 & $24.54^{+0.20}_{-0.17}$ \\
F160W & 1528 & $24.57^{+0.18}_{-0.17}$ \\
$Y_{\rm HSC}$ & 976 & $24.48^{+0.08}_{-0.07}$ \\
$Y$ & 1020 & $24.68^{+0.25}_{-0.21}$ \\
$J$ & 1248 & $24.54^{+0.25}_{-0.21}$ \\
$H$ & 1635 & $24.78^{+0.30}_{-0.25}$ \\
$K_s$ & 2144 & $24.74^{+0.32}_{-0.24}$ \\
\end{tabular}
\label{tab:totalphot} 
\end{table}

 \begin{table*}
\caption{Resolved photometry of CR7's individual components as measured with $0.5''/0.5''$/$0.8''$ (MUSE/{\it HST}/ground-based data) diameter apertures, including aperture corrections based on {\it HST} morphology. Ly$\alpha$ flux is in $10^{-17}$ erg s$^{-1}$ cm$^{-2}$. We note that care must be taken in interpreting resolved Ly$\alpha$ fluxes, as the Ly$\alpha$ emission may not be originating from the same location as the UV emission as indicated by differences in the morphology. }
\begin{tabular}{lrrrrrrrrrr}
ID & f$_{\rm Ly\alpha}$ & F110W & F140W & F160W  & $Y_{\rm HSC}$ & $Y$ & $J$ & $H$ & $K_s$ \\ \hline
A & 6.51$^{+0.16}_{-0.14}$   & $24.87^{+0.03}_{-0.03}$ & $24.96^{+0.06}_{-0.06}$ & $24.99^{+0.03}_{-0.03}$ & $25.06^{+0.08}_{-0.08}$ & $25.37^{+0.19}_{-0.17}$ & $25.01^{+0.19}_{-0.16}$ & $25.31^{+0.36}_{-0.28}$ & $25.07^{+0.23}_{-0.19}$\\

B & 0.84$^{+0.21}_{-0.21}$   & $26.97^{+0.16}_{-0.14}$ & $26.78^{+0.32}_{-0.24}$ & $26.91^{+0.16}_{-0.13}$ & $26.51^{+0.32}_{-0.24}$ &  $27.50^{+1.30}_{-0.73}$ & $26.69^{+1.00}_{-0.55}$ & $26.63^{+1.20}_{-0.65}$ & $26.61^{+1.02}_{-0.57}$\\

C & 0.77$^{+0.17}_{-0.16}$   & $26.21^{+0.09}_{-0.09}$ & $26.29^{+0.22}_{-0.24}$ & $26.17^{+0.09}_{-0.08}$ & $26.48^{+0.28}_{-0.23}$  & $26.29^{+0.54}_{-0.37}$ & $26.32^{+0.52}_{-0.35}$ & $26.29^{+0.96}_{-0.56}$ & $25.98^{+0.58}_{-0.38}$\\

\end{tabular}
\label{tab:resolvedphot}
\end{table*}
 
\subsection{Photometric model} \label{sec:photmodel} 
We describe the spectral energy distribution with a simple model that contains the Ly$\alpha$ emission line and a UV continuum that breaks below the Ly$\alpha$ wavelength due to attenuation by the IGM \citep[e.g.][]{Madau1995}, which is relevant for the $Y_{\rm HSC}$ and F110W photometry. The UV continuum is characterised by a normalisation (M$_{1500}$, the absolute UV magnitude at $\lambda_0=1500$ {\AA}) and a single power-law slope ($\beta$). This model therefore ignores additional rest-frame UV emission and absorption features. 

The best-fitted model parameters and their 68 \% confidence intervals are found by simulating a large grid of models with varying UV luminosity, UV slope and Ly$\alpha$ luminosity. Each model is shifted to $z=6.6$ and convolved with the filter transmission curves to be compared to the observed magnitudes of the ground-based and {\it HST} data. We compute the likelihood ($\mathcal{L} \propto \exp(-\chi^2/2)$) of each model by comparing the model to the observed magnitudes and their uncertainties and find the model with the highest likelihood. We note that we use the logarithmic Ly$\alpha$ luminosity in the $\chi^2$ calculation for consistency with the use of magnitudes in the other photometry data. For each of the fitted parameters, uncertainties are derived from the 16th and 84th percentiles of the marginalised posterior distribution. 

Our model results are listed in Table $\ref{tab:phot_models}$, where we list the results for CR7 as a whole and per component. We list the results obtained when including only the {\it HST} and MUSE data. We note that results are in good agreement when ground-based imaging data is also included, although the UltraVISTA data tends to drive the results to a somewhat redder UV slope due to the brightness in the $K_s$ band and relative faintness in the $Y$ band. We note that clump A-2 contributes $\approx10$ \% of the flux in component A and therefore has an absolute magnitude M$_{1500}\approx-19.4$.

\begin{table}
\caption{Best-fit values to the rest-frame UV SED model of CR7 as a whole and for its individual components using {\it HST} and MUSE data.}
\centering
\begin{tabular}{lrr}
ID & M$_{1500}$ & $\beta$  \\ \hline
Total & $-22.24^{+0.09}_{-0.09}$ & $-2.0\pm0.55$  \\
A & $-21.92^{+0.02}_{-0.03}$ & $-2.35^{+0.10}_{-0.20}$   \\
B & $-19.82^{+0.11}_{-0.13}$ & $-1.7\pm0.5$   \\
C & $-20.61^{+0.07}_{-0.08}$ & $-2.0\pm0.4$   \\

\end{tabular}
\label{tab:phot_models}
\end{table}

\subsection{The Lyman-$\alpha$ Equivalent Width} \label{sec:measure_EW}
Here we present measurements of the Ly$\alpha$ equivalent width (EW), which is the Ly$\alpha$ flux divided by the continuum flux density. While Ly$\alpha$ flux density is well measured from the MUSE data, the continuum level needs to be estimated with photometry as there is no significant coverage of wavelengths redder than Ly$\alpha$ in the MUSE data. We explore two different methods. In the first method we use the continuum level as measured in the UltraVISTA $Y$ band and extrapolate it to 1216 {\AA} assuming a flat spectral slope ($\beta=-2$). This filter covers $\lambda_0=1280-1410$ {\AA} at $z=6.6$ and is therefore the closest in wavelength to the Ly$\alpha$ line, while not including the line itself. In the second method we use the results from the photometric modelling from \S $\ref{sec:photmodel}$ using MUSE and {\it HST} data in order to estimate the continuum around Ly$\alpha$ by simultaneously modelling the UV slope. We use aperture-corrected photometry as described earlier. The results are listed in Table $\ref{tab:EW}$. 

It is not straightforward to interpret the Ly$\alpha$ EWs for the individual components. As we showed in \S $\ref{sec:lyamorph}$, the morphology of the Ly$\alpha$ emission is significantly different from that of the UV continuum emission. This indicates that the Ly$\alpha$ flux measured at the position of one of the components is not necessarily physically associated to this component. Therefore, we measure the total EW including all continuum and line emission, which is the only possible model-independent measurement. Comparing the total EW in the different methods, we measure EW$_0 = 107^{+28}_{-22}$ {\AA} when the continuum is estimated with the $Y$ band and a lower EW$_0 = 74^{+16}_{-14}$ {\AA} when the {\it HST} data is used to estimate the continuum. This is caused by a fainter $Y$ band flux compared to the best-fitted continuum model (\S $\ref{sec:photmodel}$). In principle one explanation could be broad Ly$\alpha$ absorption on top of strong, narrow emission, as for example recently observed in lower-redshift analogues \citep{Erb2019,Jaskot2019}. However, such hypothetical absorption feature would need to be broader than in these known cases in order to extend far into the $Y$ band. Whether the UV continuum is more complex around the $Y$ band or systematic offsets are present in the $Y$ band photometry can only be evaluated with future data.

We also measure the EW assuming that the ionising radiation associated to the UV clump A is responsible for the production of the majority of Ly$\alpha$ photons by using clump A's UV continuum emission and the total Ly$\alpha$ luminosity. This results in an estimated EW=$101^{+11}_{-9}$ {\AA} to EW= $200^{+42}_{-33}$ {\AA} (for the two methods). Additionally, we also show that the effect of a small correction for Ly$\alpha$ component 2 (with distinct line-profile, see \S $\ref{sec:2ndline}$), is only marginal ($\approx2$ \%; see Table $\ref{tab:EW}$). If we only associate `core-like' Ly$\alpha$ emission (\S $\ref{sec:lyamorph}$) to the UV continuum of clump A, we measure EWs that are a factor $\approx1.5$ lower. We discuss these measured EWs in \S $\ref{sec:discuss_EW}$.

Finally, we measure the EW for Ly$\alpha$ component 2, assuming it originates from the nearby UV component B and find a moderate EW$\approx20$ {\AA}. There is no significant Ly$\alpha$ emission that is distinctly observed to originate from UV component C. We derive a rough upper limit on the EW of this component by combining its UV continuum with the Ly$\alpha$ flux from component 2, which is a conservative upper limit of the Ly$\alpha$ flux we could have associated to component C. This results in an EW$<10$ {\AA}, implying little Ly$\alpha$ production or escape from this component. 
 
 \begin{table}
\caption{Rest-frame Ly$\alpha$ EW of CR7 for different scenarios. Different columns show different methods to measure EW, either using the UltraVISTA $Y$ band as continuum level around Ly$\alpha$ or using the {\it HST}-based model.}
\begin{tabular}{lrr}
Scenario  & EW$_{0, Y}$ [\AA] & EW$_{0, \rm HST}$ [\AA]   \\ \hline
Total cont. \& Total Ly$\alpha$ & $107^{+28}_{-22}$ & $74^{+16}_{-14}$   \\
A cont., Core-like Ly$\alpha$ & $141^{+29}_{-23}$  & $68^{+6}_{-6}$ \\
A cont., Total Ly$\alpha$ & $200^{+42}_{-33}$ & $101^{+11}_{-9}$  \\

A cont., Total Ly$\alpha$ of main line & $197^{+40}_{-33}$ & $99^{+11}_{-9}$  \\ 

B cont., Ly$\alpha$ component 2 & $28^{+48}_{-18}$ &  $14^{+6}_{-4}$  \\

C cont., limit  & $<9$ &  $<6$ \\  
\end{tabular} \label{tab:EW}
\end{table}

\section{A single source illuminating a complex structure} \label{sec:results}
In this section we combine our results and compare these to other studies to argue that clump A is the single prevalent powering source for the Ly$\alpha$ emission in CR7 that is making an extended gas distribution visible.
 
As illustrated in Fig. $\ref{fig:1}$, the Ly$\alpha$ emission in CR7 appears to be rather smooth, particularly in comparison to the clumpy UV continuum and the [CII] line emission at matched resolution \citep{Matthee2017ALMA}. How can such differences be explained? Using a 3D analysis, we find that the Ly$\alpha$ emission can be spectrally decomposed in a largely dominant extended component and a faint redder component whose position is close to faint components identified in UV and [CII] (Fig. $\ref{fig:MUSE_TWO_LINECOMPONENT}$). The dominant part of the Ly$\alpha$ emission peaks close from the brightest UV component and this Ly$\alpha$ component extends in the direction of the other UV components. The additional, much fainter, Ly$\alpha$ component appears to originate from UV clump B and no distinct Ly$\alpha$ component is observed around clump C.

The Ly$\alpha$ halo in CR7 appears offset by $1.3\pm0.2$ kpc from the peak of the UV emission (\S $\ref{sec:astrometry}$). Such offsets are also reported in several cases in the literature. \cite{Hoag2019} report a distribution of spatial offsets with a spread of $\approx1.2$ kpc in a sample of UV-selected galaxies at $z=4-5.5$. More directly comparable, \cite{Jiang2013_MORPH} reports qualitatively that the Ly$\alpha$ emission in bright merging systems at $z\approx6.5$ tends to be offset from the main UV component, while Ly$\alpha$ typically is co-spatial for LAEs with a single UV component. \cite{Ouchi2013} report that the Ly$\alpha$ emission in Himiko, a similar triple UV component system as CR7 at $z=6.59$, peaks close to ($\approx1$ kpc), but not exactly on top of, the brightest UV continuum component. In Himiko, the peak of the Ly$\alpha$ emission is perfectly co-spatial with a [CII] emitting component \citep{Carniani2018Himiko}, while for CR7 it is not.

The resolved line-profile fitting (\S $\ref{sec:lineprofile}$) indicates that (away from other UV components and the additional Ly$\alpha$ component) the main Ly$\alpha$ component becomes slightly redder as a function of distance from the Ly$\alpha$ centre (by $\approx30-40$ km s$^{-1}$ at a distance of $\approx3.5$ kpc; left panel of Fig. $\ref{fig:MUSE_RESOLVED_LINE}$). This resembles recent results at $z\approx3-4$ \citep{Claeyssens2019,Leclercq2020} who report somewhat redder Ly$\alpha$ lines at lower surface brightness compared to the Ly$\alpha$ line profile at peak surface brightness, and which they suggest to be indicative of resonant scattering and to support the idea that most of the Ly$\alpha$ emission originates from the UV peak.

\section{On the profile and brightness of the Ly$\alpha$ halo} \label{sec:halo_comparison}
In this section we compare the brightness, scale length and the ellipticity of the extended Ly$\alpha$ halo in CR7 to those of other LAEs studied in the literature. 

\begin{figure}
\includegraphics[width=8.6cm]{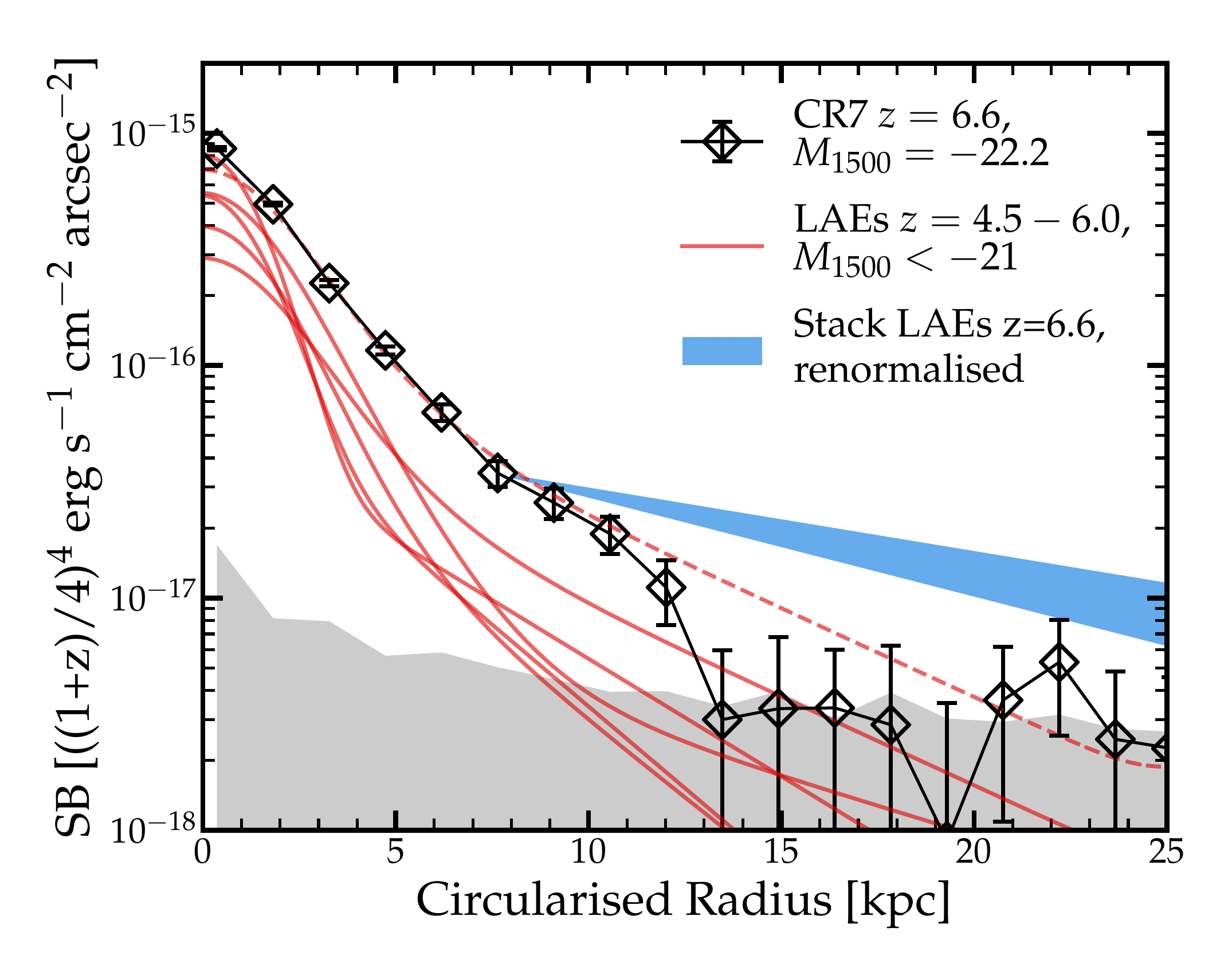}\\
\caption{Surface brightness profile of CR7's Ly$\alpha$ emission (black diamonds) extracted as detailed in Appendix $\ref{sec:1DSB}$, corrected for surface brightness dimming with respect to $z=3$. The noise level is shown in grey. The red solid lines show the best-fit Ly$\alpha$ surface brightness profiles in the five UV brightest LAEs at $z=4.5-6.0$ from \citet{Leclercq2017} (IDs 53, 1185, 1670, 6462 and 7001), for clarity of comparison convolved with the PSF of the CR7 data. The dashed line shows that the Ly$\alpha$ SB profile of ID 1185 at $z=4.5$ is remarkably similar to the one in CR7, once its total Ly$\alpha$ luminosity is rescaled to the same total Ly$\alpha$ luminosity of CR7. The blue shaded region shows the typical halo profile measured in stacks of LAEs by \citet{Momose2014} renormalised to the SB of CR7 at 7.5 kpc. }
\label{fig:MUSE_SB_COMPARISON}  
\end{figure}

For a first comparison, in Fig. $\ref{fig:MUSE_SB_COMPARISON}$ we show the (redshift dimming-corrected) 1D Ly$\alpha$ surface brightness profile of CR7 (see Appendix $\ref{sec:1DSB}$ for details) and the profiles of the five UV brightest LAEs at $z=4.5-6.0$ observed with MUSE by \cite{Leclercq2017}. These five LAEs have a typical UV luminosity of $M_{1500}=-21.1$ and Ly$\alpha$ luminosity $1.5\times10^{43}$ erg s$^{-1}$ and are thus a factor $\approx3$ fainter than CR7 while having similar Ly$\alpha$ EW. Besides this normalisation difference, the SB profile of CR7 appears quite similar to the SB of the comparison sample. This is illustrated in particular by the dashed red line in Fig. $\ref{fig:MUSE_SB_COMPARISON}$, which as an example shows that the SB profile of the MUSE LAE with ID 1185 appears extremely similar to CR7's profile, once rescaled for the luminosity difference. Other LAEs in the comparison sample have more compact core-emission compared to CR7, but this may be plausibly explained by their difference in UV luminosity and the relation between the UV size and UV luminosity \citep[e.g.][]{Shibuya2015}.

Focussing on the scale length, we find that the scale length of the exponential halo in CR7 is very similar to the typical scale length ($r_s = 3.8^{+3.1}_{-2.0}$ kpc, 68th percentiles) measured in individual Ly$\alpha$ halos of fainter galaxies at $z=3-5$ \citep{Leclercq2017}. Additionally, the fraction of the Ly$\alpha$ flux that originates from the halo component is similar to the typical halo fraction of $66\pm20$ \% in LAEs at $z=3-5$. The scale length also resembles that of the extended Ly$\alpha$ emission measured in another bright LAE at $z=6.5$ \citep{Matthee2019MUSE}. 

These relatively compact scale lengths (although still a factor $\approx10$ larger than the UV continuum) are in contrast to the stacking results from \cite{Momose2014}, who measure a significantly larger scale length of $r_s=12.6^{+3.3}_{-2.4}$ kpc in a stack of fainter LAEs at $z=6.5$. These results are illustrated by the blue shaded region in Fig. $\ref{fig:MUSE_SB_COMPARISON}$, which is renormalised to the SB of CR7 at 7.5 kpc as this is roughly the radius where the SB profile traces the halo component almost exclusively. We note that \cite{Momose2014} only fit the halo component at r $>11$ kpc ($2''$), so we have extrapolated these results slightly. The difference between the SB profile of CR7 and that measured by \cite{Momose2014} indicates that fainter LAEs have halos with larger scale length (see also \citealt{Santos2016}). However, we note that the average halo scale length measured by the stacking analysis from \cite{Momose2014} at $z\approx3$ is also significantly larger than the average scale length measured in individual halos with MUSE by \cite{Leclercq2017}, even though these systems have similar luminosity, indicating that differences in the data and the methodology may dominate the discrepancy. 

We conclude that the scale length of CR7's Ly$\alpha$ halo is rather typical at $z=4.5-6.0$, particularly for LAEs for which Ly$\alpha$ halos have been measured individually with similar methodology and instrumentation as our measurements. This indicates that the CGM around CR7 is comparable to post-reionisation galaxies. The discrepancy to the results on fainter LAEs at $z=6.6$ by \cite{Momose2014} can be resolved when individual halos in fainter LAEs at $z=6.6$ are measured with MUSE, although this will require a significant investment of observing time.

The extended Ly$\alpha$ emission around CR7 is clearly elongated, with an ellipticity of $\approx0.5$ meaning that the semi-major axis is roughly twice the semi-minor axis. This is in contrast to the low ellipticity of $\approx0.15$ measured in another UV luminous LAE at $z=6.5$ \citep{Matthee2019MUSE}. Most earlier studies of extended Ly$\alpha$ emission impose circular symmetry \citep[e.g.][]{Steidel2011,Momose2014,Leclercq2017} motivated by visual inspection. \cite{Wisotzki2015} also impose circular symmetry, but confirm that their results are not influenced significantly by this assumption. \cite{Wisotzki2015} mention that roughly 75 \% of their objects have axis ratios higher than 0.5 (ellipticity $<0.5$) and displacements $<0.2''$. This indicates that the elongated shape of the Ly$\alpha$ halo of CR7 is not very uncommon. 

One possible explanation for the elongated shape of CR7's Ly$\alpha$ halo is that CR7 is a multiple component galaxy, possibly due to a merger event or coeval clumps of star formation. This is unlike the LAEs in the sample from \cite{Wisotzki2015} that appear as single component systems. As Fig. $\ref{fig:1}$ shows, the Ly$\alpha$ emission from CR7 is preferentially extended in the direction of fainter UV clumps, particularly the faintest UV components (clumps A-2 and B). As shown in \S $\ref{sec:2ndline}$, $\approx98$ \% of the Ly$\alpha$ emission is observed with a line-profile that is similar in shape to the line profile at the peak position of the Ly$\alpha$ emission close to the brightest UV peak. This indicates that no significant amounts of Ly$\alpha$ photons originate from these fainter UV components. Therefore, it is more likely that the elongated shape of the Ly$\alpha$ halo is caused by the distribution of hydrogen gas that extends in the direction of the other UV clumps rather than being caused by multiple production sites of Ly$\alpha$ photons. 

We note that the extended Ly$\alpha$ emission around Himiko also appears to be elongated along the direction where multiple UV components are seen \citep{Ouchi2013}, suggesting this could be a common scenario among bright LAEs that consist of multiple components.

\section{What is the powering source of the Ly$\alpha$ emission?} \label{sec:discussion}
In this section we combine the various measurements presented previously with earlier observations of CR7 (in particular ALMA observations and rest-frame UV spectroscopy; \citealt{Matthee2017ALMA,Sobral2019}) to discuss what is the powering source of the high Ly$\alpha$ luminosity. In particular, we focus our discussion on distinguishing between Ly$\alpha$ emission that originates as recombination radiation powered by either young stars or an AGN.

\subsection{On the high EW for CR7's Ly$\alpha$ line profile}  \label{sec:discuss_EW}
Here we estimate the intrinsic Ly$\alpha$ EW of CR7 in order to address whether star formation from clump A alone can power the Ly$\alpha$ emission. The {\it observed} Ly$\alpha$ EW is related to the amount of Ly$\alpha$ photons that are produced and that are not destroyed by dust and furthermore not scattered by IGM gas intervening along our line of sight, relative to the UV continuum \citep[e.g.][]{SM2019}. As such, the Ly$\alpha$ EW increases with increasing ionising photon production efficiency at fixed escape fraction (\citealt{Maseda2020}; related predominantly to age of the stellar populations, but also to metallicity, binary fraction and the shape of the initial mass function; IMF). Ly$\alpha$ EW decreases with increasing dust attenuation \citep{Matthee2016}, particularly if a high column density of neutral hydrogen leads to higher travelled path lengths of Ly$\alpha$ photons compared to UV continuum photons due to resonant scattering \citep[e.g.][]{Scarlata2009,Henry2015}. 

While it is challenging to directly interpret observed Ly$\alpha$ EWs without further information such as the H$\alpha$ luminosity or the dust attenuation, it is possible to use the observed EW to broadly address the nature of the ionising source. In particular, as discussed in e.g. \cite{CharlotFall1993,Raiter2010}, `normal' star formation (i.e. Population II stars with a standard IMF) is expected to produce a maximum Ly$\alpha$ EW of $\approx240$ {\AA}. A higher Ly$\alpha$ EW likely requires additional or more extreme sources of ionising photons such as a nearby AGN \citep[e.g.][]{Marino2018}, very young starbursts \citep{Maseda2020} or exotic stellar populations \citep{Raiter2010}. 

\cite{Sobral2015} measured a total Ly$\alpha$ EW$_0=211\pm20$ {\AA} for CR7 based on a combination of narrow-band imaging and spectroscopy. This would place CR7 in the regime of extreme stellar populations and/or an AGN contribution, particularly as it is likely that we are not observing the total intrinsic EW due to dust absorption in the ISM/CGM and scattering in the IGM. The estimate by \cite{Sobral2015} is based on shallower $Y$ band photometry and a Ly$\alpha$ flux that is a factor $\approx1.5$ higher compared to the MUSE measurement. This can be attributed to an over-correction for the filter transmission at the wavelength at which CR7's Ly$\alpha$ line is detected in the NB921 filter in \cite{Sobral2015}. Note that despite our new and lower MUSE-based Ly$\alpha$ flux, we can still recover an EW as high as that reported in \cite{Sobral2015}, but only if adopting the $Y$ band photometry (which is only marginally consistent with the other broad-band photometry; Table $\ref{tab:EW}$). The Ly$\alpha$ EW that is estimated with a combination of the MUSE data and the full multi-band {\it HST} photometry indicates a lower observed EW$_0$ of $99^{+11}_{-9}$ {\AA} assuming that the vast majority of Ly$\alpha$ flux originates from clump A (see \S $\ref{sec:measure_EW}$), which we consider in our discussion below.

\begin{figure}
\includegraphics[width=8.6cm]{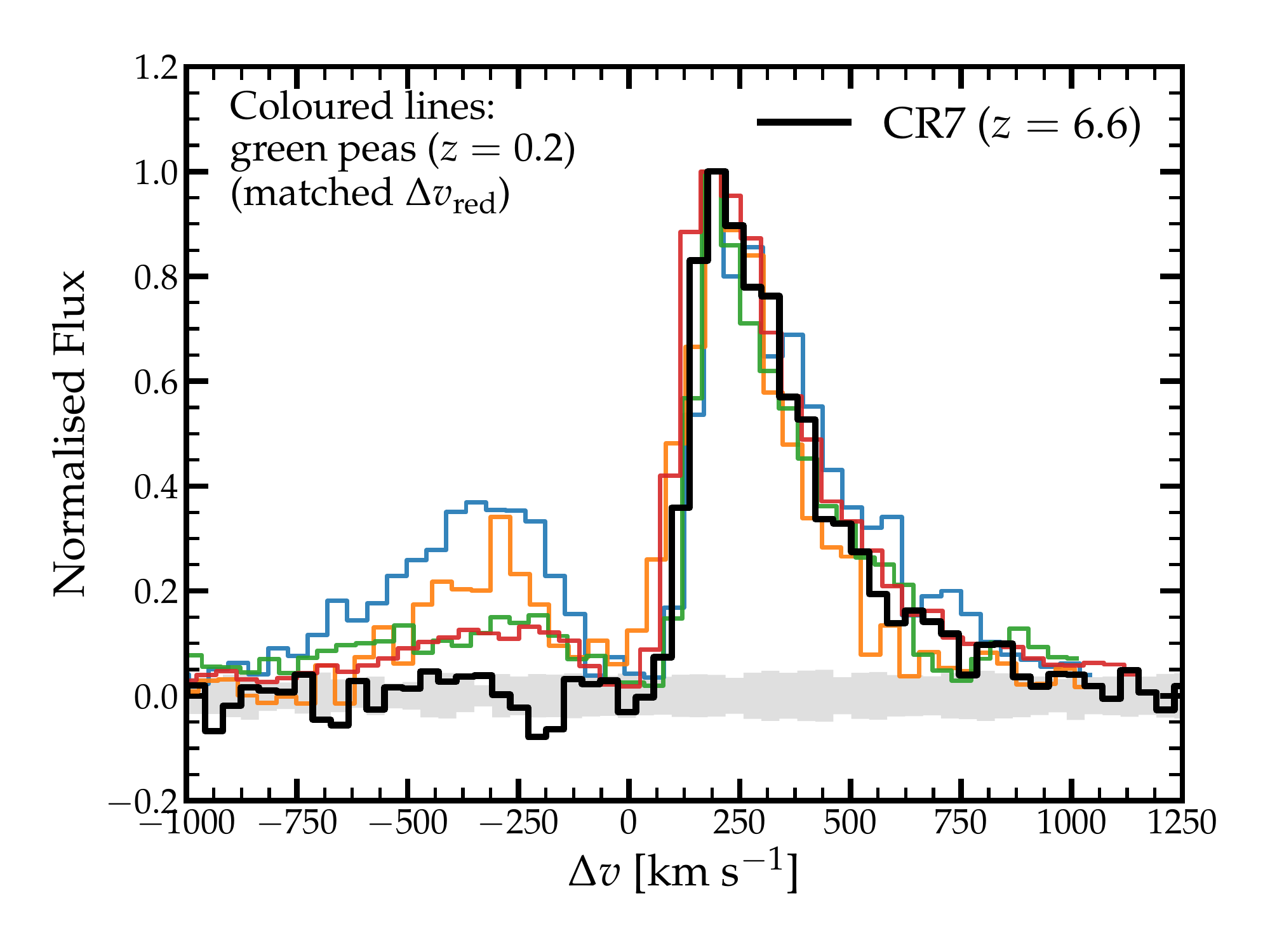}\\
\caption{Observed Ly$\alpha$ profiles of CR7 (thick black line, from MUSE data) and four green pea galaxies at $z\approx0.2$ that are selected to be matched in $\Delta v_{\rm red}$ (coloured lines). Ly$\alpha$ profiles are normalised to the peak flux in the red part of the line. The grey shaded area shows the 1$\sigma$ noise level in the MUSE data. The S/N and resolution of the MUSE data of CR7 at $z=6.6$ are virtually indistinguishable to the observations of the green pea galaxies in the red part of the spectrum. This may be expected if they are intrinsically similar systems attenuated by (redshift dependent) IGM absorption in the blue. The Ly$\alpha$ profiles from the green pea galaxies are from {\it HST}/COS observations of J1137+3524, J1054+5238, J1018+4106 and J0822+2241. These profiles are adapted from \citet{Yang2017}. }
\label{fig:comparison_Yang}
\end{figure}

It is likely that a low IGM transmission impacts the observed Ly$\alpha$ EW, particularly at $z>6$. The simulation by \cite{Laursen2011} suggests that on average sight-lines at $z=6.6$ the IGM transmission jumps from $\approx0$ \% at $v< +100$ km s$^{-1}$ (where $v$ is the velocity with respect to the systemic), to $\approx 100$ \% at velocities higher than 100 km s$^{-1}$. It is therefore likely that we are missing the entire blue part of the line, implying that the observed EW is lower than the intrinsic one (but see \citealt{Matthee2018} for a rare counter example at $z=6.59$). On the other hand, as almost all of CR7's Ly$\alpha$ photons are observed at $> +200$ km s$^{-1}$ with respect to the systemic (Fig. $\ref{fig:comparison_Yang}$)\footnote{Note that the spectral resolution of FWHM=70 km s$^{-1}$ smears a fraction of photons to the blue in intrinsically asymmetric profiles making them artificially appear at $v<200$ km s$^{-1}$.}, it is likely that the Ly$\alpha$ escape fraction of the Ly$\alpha$ photons that emerge on the red side of the systemic is mostly set by the ISM conditions in CR7.

To obtain an estimate of the intrinsic Ly$\alpha$ EW$_{int}$ (as opposed to the observed, rest-frame EW that we denote with EW$_0$), we compare the shape of CR7's Ly$\alpha$ line to the Ly$\alpha$ properties of green pea galaxies (GPs), which are often considered analogues of high-redshift galaxies, with high quality rest-frame UV and optical spectroscopy \citep[e.g.][]{Amorin2010,Henry2015}. As these galaxies are typically at $z\approx0.2-0.3$, it is reasonable to assume the IGM transmission of Ly$\alpha$ photons is 100 \%, even on the blue side of the systemic redshift. In particular, we select the four GPs from the sample by \cite{Yang2017} that are closest to CR7 in terms of $\Delta v_{\rm red}$, the peak velocity of the red part of the Ly$\alpha$ line. We do not impose additional criteria. We note that the UV and Ly$\alpha$ luminosities of these galaxies are roughly one order of magnitude lower than those of CR7 (M$_{1500} =-19.4$ to $-21.2$, L$_{\rm Ly \alpha} = 1-3 \times 10^{42}$ erg s$^{-1}$). These galaxies have gas-phase metallicities 12+log(O/H)$\approx8.0$ and specific SFRs $\approx5$ Gyr$^{-1}$, which is typical for galaxies at $z\sim7$ \citep{Stark2013}. Visually these galaxies appear to be dominated by a single bright clump in the rest-frame UV, but several show a secondary fainter component on $\sim1$ kpc distance \citep{Yang2017}.

The Ly$\alpha$ profiles of these GPs and the Ly$\alpha$ line from CR7 (integrated over the total system\footnote{The results would be unchanged if we would use the Ly$\alpha$ spectrum extracted at the peak position.}) are shown in Fig. $\ref{fig:comparison_Yang}$. It is remarkable that, in addition to the velocity offset, the width and the shape of the red line are also well matched.\footnote{We note that the spectral resolution of the Ly$\alpha$ observations of GPs is not fully known due to the unknown extent of the Ly$\alpha$ lines in the {\it HST}/COS aperture. However, \cite{Orlitova2018} estimate a resolution FWHM of $\approx100$ km s$^{-1}$, which is comparable to the MUSE data of CR7 at $z=6.6$.} This is likely a consequence of a correlation between the Ly$\alpha$ line width and the velocity shift that is well understood in resonant scattering models with simple geometries \citep[e.g.][]{Neufeld1991,Verhamme2018}. This result implies that the effective HI column densities (i.e. the sightline-averaged column density through which the observed Ly$\alpha$ photons scattered) in the ISM of CR7 is similar to that in the GPs with similar velocity shifts. It also supports the use of GPs as analogues of high-redshift galaxies.

A clear difference between CR7 and the GPs is that all low-redshift Ly$\alpha$ profiles show emission on the blue side of the systemic that contains between 16-33 \% of the total Ly$\alpha$ flux (25 \% on average), while this is not seen in CR7. This is likely due to the impact of the IGM at $z=6.6$, as discussed above, and implies that the red part of the Ly$\alpha$ line is not significantly affected by the IGM. Correcting for the average fraction of missing blue Ly$\alpha$ photons would result in a Ly$\alpha$ EW$_{0}\approx 130$ {\AA} for CR7. In the GPs, the separations of the two Ly$\alpha$ peaks range from 250 to 520 km s$^{-1}$ and, on average, suggest an escape fraction of ionising photons of $\approx 2$ \% \citep{Verhamme2015,Izotov2018}.

The most important difference between CR7 and the GPs is that the Ly$\alpha$ EWs of the GPs are significantly smaller (EW$_{0, {\rm red}} = 25\pm5$ {\AA}; here we only consider the flux in the red part of the line for a fair comparison). \cite{Jaskot2019} also report similarly low EW$_0$ measurements in their sample of GPs with $\Delta v_{\rm red} \approx +200$ km s$^{-1}$. The EW difference could either be explained by a lower relative dust attenuation of Ly$\alpha$ photons compared to the UV continuum (i.e. a higher ratio of Ly$\alpha$ escape fraction to UV escape fraction) and/or a higher intrinsic Ly$\alpha$ EW in CR7 compared to the GPs. 

We estimate the intrinsic Ly$\alpha$ EW$_{int, {\rm tot}}$ (including blue photons) of the GPs as follows:
\begin{equation}
EW_{int, {\rm tot}} = EW_{0, {\rm red}} \times \frac{f_{\rm esc, continuum}}{f_{\rm esc, Ly\alpha, red}},
\end{equation}
where $f_{\rm esc, Ly\alpha, red}$, the Ly$\alpha$ escape fraction on the red side of the systemic, accounts for the attenuated Ly$\alpha$ photons and $f_{\rm esc, continuum}$ accounts for attenuation in the UV continuum level due to dust. Assuming a \cite{Calzetti2000} dust attenuation curve, this attenuation is proportional to $f_{\rm esc, continuum}= 10^{-0.4 \kappa E(B-V)}$ where $\kappa=12$ at the Ly$\alpha$ wavelength and where $E(B-V)$ is estimated from the Balmer decrement (see \citealt{Yang2017}). The measured $E(B-V)$ range from 0.04 to 0.20 and $f_{\rm esc, Ly\alpha}$ ranges from $4 - 15$ \%.\footnote{The majority of Ly$\alpha$ photons escape on the red side of the systemic, resulting in Ly$\alpha$ escape fractions of the four green peas ranging from 3-13 \% when only red photons are considered.} As a result, we estimate a mean intrinsic EW$_{int, {\rm tot}}$ of 135 {\AA} for the GPs (ranging from 75 to 200 {\AA}), typically a factor $\approx4$ higher than EW$_0$. We find that the estimated intrinsic Ly$\alpha$ EW correlates well with the H$\alpha$ EW, following EW$_{int, \rm tot, Ly\alpha} \propto $ 1/4 EW$_{0, \rm H\alpha}$. 

Now, we assume that the difference between the intrinsic and observed EW is similar in CR7, motivated by the fact that the (red parts) of the Ly$\alpha$ profiles of CR7 and the GPs are matched and therefore that the path length of Ly$\alpha$ photons relative to UV photons may be similar, which {\it at fixed dust content}, would imply similar relative attenuation and EW correction \citep[e.g.][]{Scarlata2009}. This assumption would imply that CR7 has an intrinsic Ly$\alpha$ EW$_{int} \approx 500$ {\AA} and H$\alpha$ EW$_0\approx2000$ {\AA}. Such high intrinsic EW could be powered by a relatively young and low metallicity  ($\lesssim3\times10^7$ yr, $Z\lesssim 0.004$) starburst \citep{Maseda2020}. At $z\sim4-5$, such high H$\alpha$ EWs are only observed in very faint LAEs with M$_{1500} \approx-18$ \citep{Lam2019}. Alternatively, the EW could also be elevated in CR7 compared to the GPs if there is less dust attenuation, which affects Ly$\alpha$ more than the continuum. Future observations of the H$\alpha$ EW and dust attenuation through the Balmer decrement could fully distinguish between these scenarios.

\subsection{Comparison to LABs and hidden AGN} \label{sec:discuss_LABs}
Besides the flux, the morphology of the Ly$\alpha$ emission could also provide insights into the powering source of the Ly$\alpha$ emission. As discussed in \S $\ref{sec:astrometry}$, it appears that the Ly$\alpha$ emission does not peak exactly on the position of the brightest UV component, but slightly towards the north-west. This somewhat resembles well-known Ly$\alpha$ blobs (LABs) at $z\approx2-3$ \citep[e.g.][]{Steidel2000,Nilsson2006}, where the peak of Ly$\alpha$ typically does not coincide with a bright Lyman-break galaxy. CR7 has a similar Ly$\alpha$ luminosity as LABs and is only slightly less extended (30 kpc, versus the typical 50 kpc) even though surface brightness dimming at $z=6.6$ is significant compared to $z=3$. The separation between the brightest UV component and the Ly$\alpha$ peak in CR7 is however significantly smaller than those reported in typical LABs. 

Based on a multi-wavelength study, \cite{Overzier2013} argue that it is likely that the majority of LABs at $z\approx2-3$ are powered by an AGN. Such AGN can be heavily obscured by dust and therefore not seen in the rest-frame UV. This is for example seen in LAB1, where there is a bright (1 mJy) sub-mm source at the position of the Ly$\alpha$ peak that is not seen in the $R$ band with limiting $R>25.7$ \citep{Steidel2000,Geach2016}. Alternatively, several luminous hot dust-obscured galaxies that are powered by AGN are observed to emit excess blue light, mimicking the colours of a typical Lyman-break galaxy \citep[e.g.][]{Eisenhardt2012,Assef2016}. 

How do such systems compare to CR7? Dust continuum emission is not detected in CR7 at $\lambda_0=160 \mu$m \citep{Matthee2017ALMA} with a limiting flux density $<7 \mu$Jy, which results in an observed IR to UV flux density ratio $\nu_{160\mu m} f_{160 \mu m} / \nu_{1500 \AA} f_{1500 \AA} < 0.02$. This limit is significantly lower than the central obscured source in LAB1, for which we estimate a limiting ratio $>4$. The typical observed IR to UV flux density ratio in blue-excess hot dust obscured galaxies is $\approx4$, similar to the limit for LAB1. If we would only take the faint UV flux of component A-2 as an extreme scenario, the limiting flux density ratio is still $\lesssim0.2$. This indicates that the IR flux density in CR7 is at least a factor 10-20 lower than expected in case the Ly$\alpha$ emission is powered by an obscured AGN, implying this scenario is unlikely. 

Besides the faintness in the IR, we also note that the Ly$\alpha$ line in CR7 is significantly narrower than the Ly$\alpha$ lines typically seen in LABs \citep[e.g.][]{Overzier2013,Sobral2018}, which can have FWHM$\approx1000$ km s$^{-1}$ at peak surface brightness \citep[e.g.][]{Herenz2020}. Finally, we note that while CR7's Ly$\alpha$ SB (corrected for redshift dimming) at a radius of 10 kpc is comparable to the Ly$\alpha$ SB around bright quasars at $z\approx3$ \citep{Borisova2016}, the profile of the extended emission around CR7 is steeper by a factor $\approx10$, indicating differences either in the powering mechanism or in the gas distribution. 

Combining the results from the Ly$\alpha$ SB profile, the Ly$\alpha$ line-profile and the UV to IR flux density ratios with earlier results from resolved rest-frame UV spectroscopy (see \S $\ref{sec:summary_previous}$), we conclude that all observational data points towards a scenario where the main powering mechanism in CR7 is a young, metal-poor starburst.

 \section{Summary} \label{sec:conclusions}
In this paper we presented sensitive spatially resolved spectroscopy of the Ly$\alpha$ emission surrounding the bright galaxy CR7 at $z=6.6$ with VLT/MUSE, combined with the most recent near-infrared imaging data from {\it HST}/WFC3 and UltraVISTA. These data allow us to measure the Ly$\alpha$ and rest-frame UV continuum morphologies, to identify additional Ly$\alpha$ emitting components with a distinct line profile and to accurately measure the total Ly$\alpha$ flux and equivalent width. We use these measurements (combined with archival sub-mm data from ALMA) to investigate what is the most likely powering source of the Ly$\alpha$ emission by comparing to various other classes of objects such as low-redshift analogues of high-redshift galaxies and obscured AGN in Ly$\alpha$ blobs.

Our main results are the following:
\begin{itemize}
\item The MUSE data confirm the bright Ly$\alpha$ emission line in CR7 with S/N$>50$ in 4 hours of integration time. The Ly$\alpha$ emission is well-resolved and extends over the three main UV emitting components of the galaxy. The total Ly$\alpha$ luminosity of $(5.3\pm0.1)\times10^{43}$ erg s$^{-1}$ is a factor 1.5 fainter than reported earlier, possibly due to uncertainties in previous narrow-band based flux measurements. Our rest-frame UV photometry based on new and deeper data is consistent with earlier measurements within the 14-86 \% percentiles and shows a blue main bright component A with M$_{1500}=-21.92^{+0.02}_{-0.03}$ and $\beta=-2.35^{+0.10}_{-0.20}$ surrounded by two fainter components (B and C) with M$_{1500}=-19.8, -20.6$, respectively and poorly constrained UV slopes $\beta\approx-2.0\pm0.5$. Our new analysis of {\it HST} data indicates a faint previously unseen component (`A-2') close to the geometric centre of the system and only $\approx1$ kpc away from clump A.

\item While CR7 is very clumpy in the rest-frame UV, the extended Ly$\alpha$ emission appears rather smooth. We show that the Ly$\alpha$ morphology can be well described by a two-component UV core + extended halo profile with a halo scale length $r_s=3.0\pm0.3$ kpc and is mostly associated to the main UV component A. The peak of the extended Ly$\alpha$ emission is somewhat offset ($\approx 1$ kpc) from the peak of the UV emission in the direction of the fainter UV components A-2 and B.

\item Using a spatially resolved line-fitting analysis, we show that the majority ($\approx98$ \%) of the Ly$\alpha$ light is well described by a single strongly skewed gaussian that appears to be slightly broader and more redshifted at the edges of the halo compared to the position of peak emission. We also identify additional faint Ly$\alpha$ emission that spatially coincides with UV clump B and that is redshifted by about $200$ km s$^{-1}$ with respect to the main emission. The contribution of this component to the total Ly$\alpha$ flux is almost negligible and it does not impact the overall morphology. 

\item The extended Ly$\alpha$ emission is strongly elongated with a major axis that is twice the minor axis. The major axis is aligned with the axis along which the multiple UV components are found. We argue that this likely traces the underlying gas distribution and is not caused by multiple separated sources of Ly$\alpha$ emission. 

\item The surface brightness profile of CR7's Ly$\alpha$ emission is similar to that observed in relatively bright LAEs at $z=4-5$. The scale length of the Ly$\alpha$ halo is significantly smaller than scale lengths measured at $z=6.6$ from stacking of fainter LAEs. This indicates that the CGM around CR7 is comparable to post-reionisation LAEs. 

\item We combine the rest-frame UV photometry with the Ly$\alpha$ luminosity to measure the EW of CR7. We note that the EW is strongly sensitive to choices in which way the continuum luminosity is estimated (due to possible systematic differences between data-sets). As the dominant part of the Ly$\alpha$ emission most likely originates from clump A, we use its {\it HST} photometry combined with the MUSE data as our best measurement and find an EW$_{0}=101^{+11}_{-9}$ {\AA} for clump A (not corrected for IGM absorption). The EW decreases to 74 {\AA} if other continuum components are included. 

\item The detailed shape of the Ly$\alpha$ line is compared to the Ly$\alpha$ lines in four low-redshift analogues of high-redshift galaxies. These are selected to have a matched peak velocity of the red part of the Ly$\alpha$ line. As a result, the full profiles of the red Ly$\alpha$ lines of these analogues resemble CR7's Ly$\alpha$ line remarkably well, indicating similar radiative transfer effects in the ISM in systems at widely different redshift, luminosity and possibly different stellar populations. On the contrary, the blue part of the Ly$\alpha$ lines that is seen in the analogues is not seen in CR7, indicating significant IGM attenuation. A significant difference is that the Ly$\alpha$ EWs in the analogues are on average a factor 4 lower than the Ly$\alpha$ EW in CR7. This indicates either a younger and/or more metal poor starburst in CR7 compared to the analogues, or a smaller enhancement of Ly$\alpha$ dust attenuation compared to the UV continuum, or both. Regardless, in all cases CR7's Ly$\alpha$ EW can be explained by a young metal-poor starburst and does not require additional ionising sources.

\item Further indications against the need for a contribution from an AGN are seen when comparing CR7 to Ly$\alpha$ blobs and other Ly$\alpha$ emitting dust-obscured AGN at $z\approx2-3$. While these systems have similar Ly$\alpha$ luminosities and also show extended Ly$\alpha$ emission, they have much broader Ly$\alpha$ lines and typically host a dust-obscured object with an IR to UV flux ratio that is at least a factor 10 higher than the upper limit in CR7.
\end{itemize}

Thus, we conclude that while the Ly$\alpha$ emission of CR7 is extremely luminous, its detailed properties such as the scale length of the Ly$\alpha$ halo, the spectral line-profile and the EW resemble more common LAEs at lower redshifts very well. The vast majority of Ly$\alpha$ emission appears to be powered by the brightest UV component. The relation between the elongation of the Ly$\alpha$ halo and the presence of the other UV components suggests that we are seeing an extended gaseous environment nurturing several star forming regions. Significant AGN contribution or extreme stellar populations are not required to power the Ly$\alpha$ emission. CR7 is likely powered by a young metal poor starburst with properties typical of much fainter galaxies. As such, it achieves a high Ly$\alpha$ luminosity and EW before the onset of significant dust attenuation as opposed to what is seen in typical star-bursting sub-mm galaxies at high-redshift. In the future, deep spatially resolved rest-frame UV and optical spectroscopy with the {\it James Webb Space Telescope} and the Extremely Large Telescopes will be able to reveal the properties of the gas and stars in this system in exquisite detail.

\section*{Acknowledgments}
We thank the referee for their constructive comments that have helped to present the data and the results in a clear way. Based on observations obtained with the Very Large Telescope, programs 0102.A-0448 and 0103.A-0272. Based on data products from observations made with ESO Telescopes at the La Silla Paranal Observatory under ESO programme ID 179.A-2005 and on data products produced by CALET and the Cambridge Astronomy Survey Unit on behalf of the UltraVISTA consortium. Based on observations made with the NASA/ESA Hubble Space Telescope, obtained at the Space Telescope Science Institute, which is operated by the Association of Universities for Research in Astronomy, Inc., under NASA contract NAS 5-26555. These observations are associated with programs \#12578, 14495, 14596. 
GP and SC gratefully acknowledge support from Swiss National Science Foundation grant PP00P2\_163824.
TN acknowledges support from the Nederlandse Organisatie voor Wetenschappelijk Onderzoek (NWO) top grant TOP1.16.057 and the Australian Research Council Laureate Fellowship FL180100060.
We have greatly benefited from the {\sc Python} programming language and particularly acknowledge the {\sc numpy, matplotlib, scipy, lmfit} \citep{Scipy,Hunter2007,Numpy} and {\sc astropy} \citep{Astropy} packages, the astronomical imaging tools {\sc SExtractor, Swarp} and {\sc Scamp} \citep{Bertin1996,Bertin2006,Bertin2010}, the morphological fitting tool {\sc imfit} \citep{Erwin2015} and the {\sc Topcat} analysis tool \citep{Topcat}.

\section*{Data availability}
The data underlying this article were accessed from the ESO and STScI archives. The raw ESO data can be accessed through http://archive.eso.org/cms.html as part of program IDs 0102.A-0448 and 0103.A-0272. The raw STScI data can be accessed through https://mast.stsci.edu as part of program IDs 12578, 14495 and 14596. The derived data generated in this research will be shared on reasonable request to the corresponding author.




\bibliographystyle{mnras}

\bibliography{bib_LAEevo.bib}




\appendix

\section{Pseudo-2D slit spectra} \label{sec:pseudoslit}
\begin{figure*}
\includegraphics[width=15.3cm]{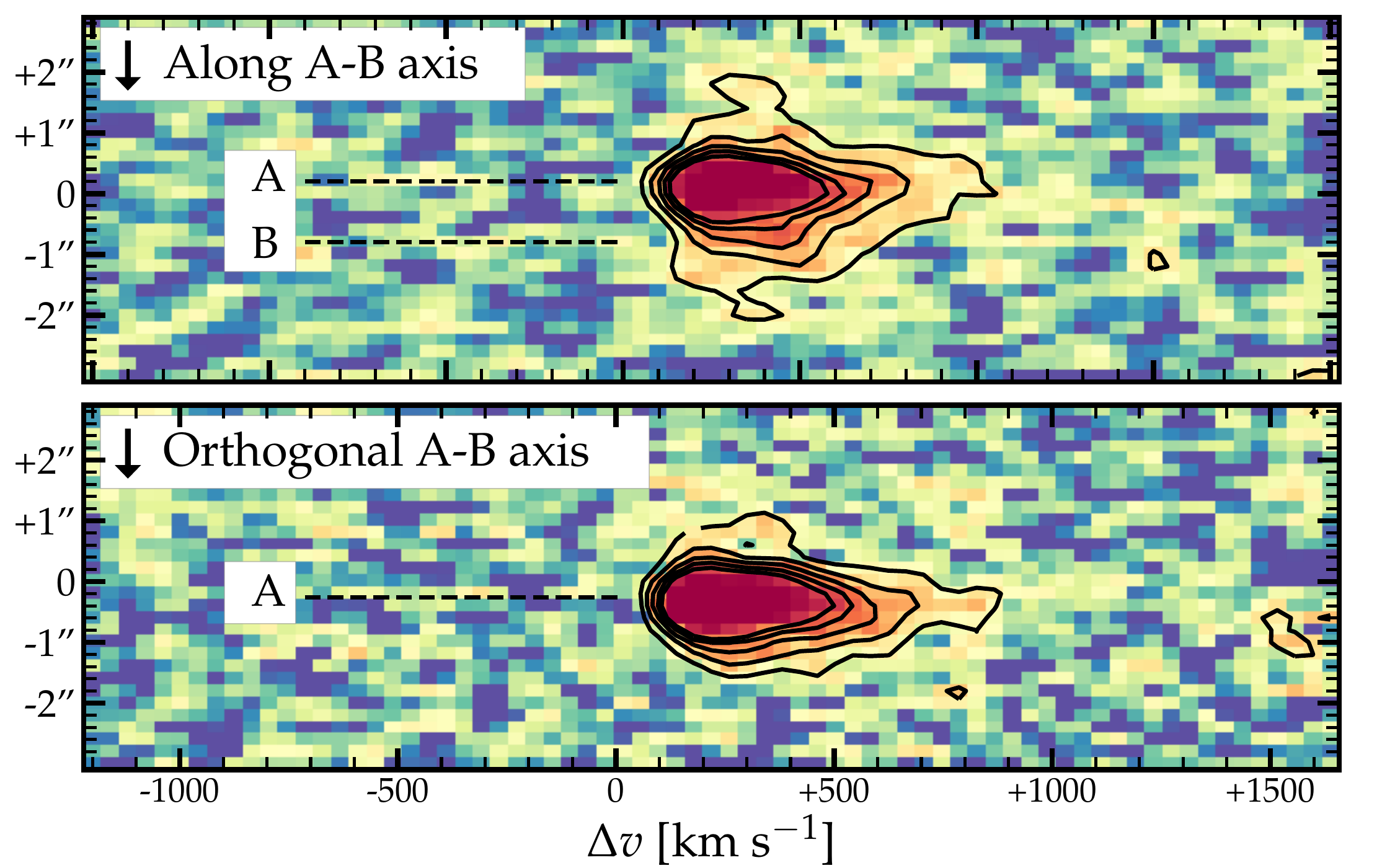} \\
\includegraphics[width=15.3cm]{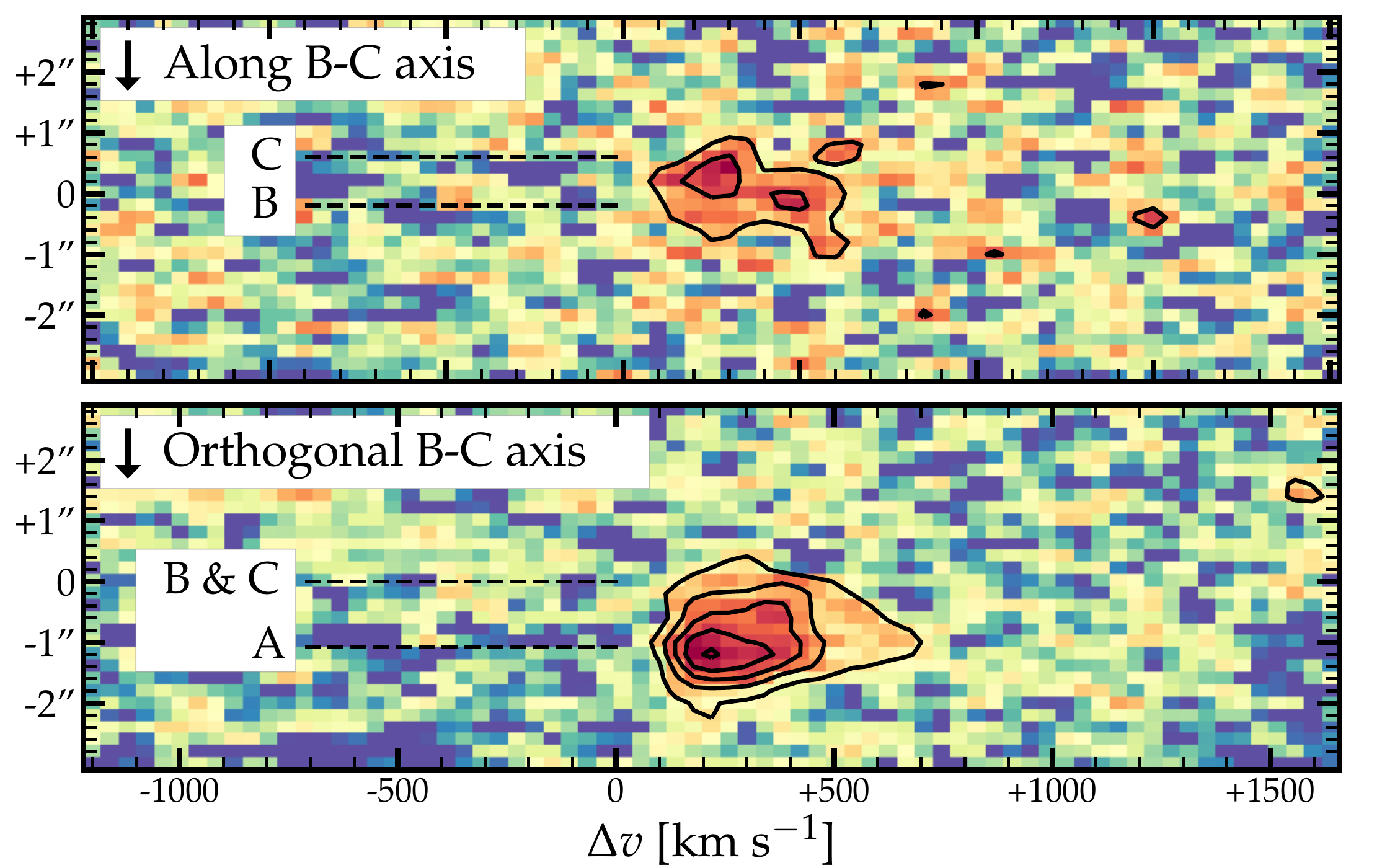} \\
\caption{Pseudo-2D slit spectra of CR7's Ly$\alpha$ line extracted from the IFU data. The slit widths are 0.6$''$. The 2D spectra are convolved with a running median filter of 2x2 pixels (corresponding to 0.4$''$ by 80 km s$^{-1}$) to increase the S/N. The colour coding follows an asinh scale to highlight emission at lower surface brightness. The contours are drawn at linearly increasing values, with the outer contour corresponding to the 3$\sigma$ level. The contours and colour coding are at the same scale in all rows. We highlight the spatial positions of components A, B and C with horizontal dashed lines. In the first two rows we show the 2D spectra along and orthogonal to the A-B axis. This pseudo-slit is centred on UV component A and has a position angle of 133 degrees. The last two rows show the 2D spectra extracted along and orthogonal to the axis of components B and C, with a position angle of 250 degrees and centred in between the clumps.   }
\label{fig:pseudoslits} 
\end{figure*}

In Fig. $\ref{fig:pseudoslits}$ we show several pseudo-2D slit spectra extracted from the MUSE IFU data. The benefit of these 2D spectra (that are sometimes called position-velocity diagrams) is that they visualise the spatial variations in the Ly$\alpha$ profile in a model-independent way (c.f. \S $\ref{sec:lineprofile}$ and \S $\ref{sec:2ndline}$) and without being limited to extractions at specific positions (e.g. Fig. $\ref{fig:1}$). 

The 2D spectra are extracted by averaging over one spatial direction within a rectangular slit with a width 0.6$''$. The 2D spectra are convolved with a running median filter of 2x2 pixels which corresponds to a 0.4$''$ by 80 km s$^{-1}$ kernel. We use this kernel to increase the S/N while simultaneously not smoothing out the detailed structures in the line-profile. The centres and position angles of the slit extractions are chosen to visualise the additional Ly$\alpha$ emitting component identified in \S $\ref{sec:2ndline}$. 

In the top two rows in Fig. $\ref{fig:pseudoslits}$ we show the extractions along and orthogonal to the A-B axis (see Fig. $\ref{fig:HSTmorph}$). The slit is centred on component A and has a position angle of 133 degrees along the A-B axis. It can clearly be seen that Ly$\alpha$ emission is more extended along the A-B axis than it is orthogonal to this axis, which visualises the elongation of the Ly$\alpha$ halo discussed in the main text. In general, the Ly$\alpha$ profile seems characterised by a single skewed gaussian profile.
In the top row, it can also be seen that there are hints of additional Ly$\alpha$ emission that peaks around the spatial position of component B and around $+400$ km s$^{-1}$. This is not seen in the slit on the second row of Fig. $\ref{fig:pseudoslits}$, consistent with the analysis in \S $\ref{sec:2ndline}$ that suggests the additional component is located around clump B.

In the bottom two rows in Fig. $\ref{fig:pseudoslits}$ we show extractions along and orthogonal to the B-C axis. The centre of the slit is located between the two components and the position angle is 250 degrees. As Ly$\alpha$ emission is fainter at this position, the S/N is lower compared to the top two rows. The third row shows that the secondary Ly$\alpha$ component around clump B can now clearly be identified at $\approx+400$ km s$^{-1}$. The emission between B and C peaks at almost the same velocity as the emission at component A and is thus extended emission from the Ly$\alpha$ halo. The bottom row further shows that the additional component does not seem to peak exactly on component B, but slightly more towards the peak of the Ly$\alpha$ emission, which is consistent with the results from \S $\ref{sec:2ndline}$.

\section{1D Surface brightness profile} \label{sec:1DSB} 
Besides modelling the Ly$\alpha$ emission in 2D, we also measure the one-dimensional surface brightness profile and fit this profile with a core and a halo component. This measurement acts particularly as a consistency check, but is also more comparable to several literature studies of extended Ly$\alpha$ emission \citep[e.g.][]{Steidel2011,Momose2014,Wisotzki2015}. 

As motivated by the Ly$\alpha$ morphology and the 2D results, we extract the one-dimensional surface brightness profile using ellipsoidal annuli. The annuli are centred on the central Ly$\alpha$ position identified in a single exponential component fit to the Ly$\alpha$ morphology (\S $\ref{sec:lyamorph}$; i.e. the red point in Fig. $\ref{fig:WCS}$), have an ellipticity 0.46 and position angle 127 degrees. We estimate the noise in each annulus from the standard deviation of the surface brightness measured in similar annuli placed in the empty sky positions identified in \S $\ref{sec:data}$. Simultaneously, these annuli are used to check that the background is subtracted properly. By estimating the errors this way we empirically account for correlated noise and do not rely on the noise image as propagated by the data reduction pipeline (which was used in the 2D modelling). We note that the number of empty apertures is limited by the difficulty in obtaining totally empty sky regions in increasingly large radii.

\begin{figure}
\includegraphics[width=8.6cm]{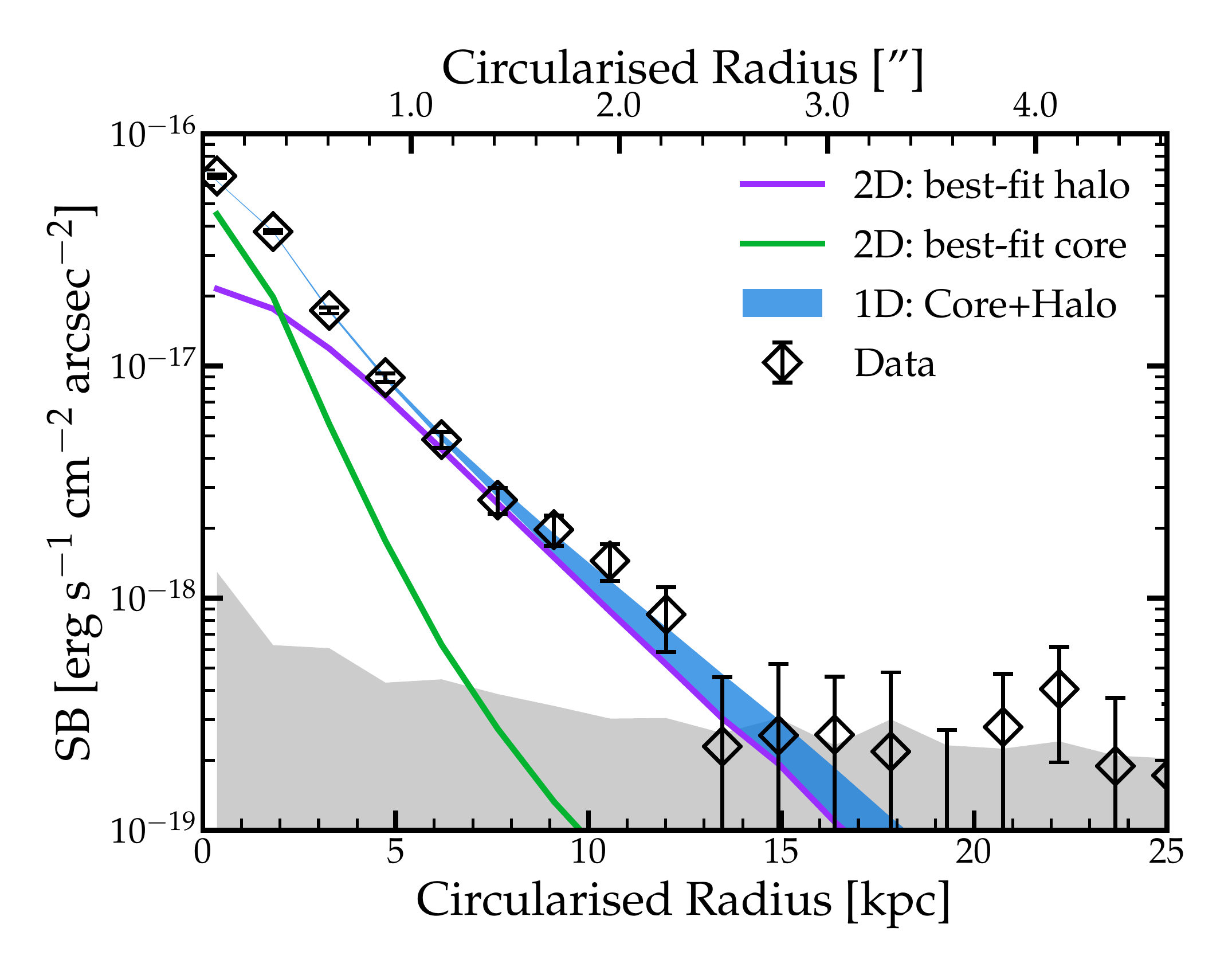}\\
\caption{Surface brightness profile of CR7's continuum-subtracted Ly$\alpha$ emission (black diamonds) extracted in elliptical annuli with ellipticity 0.46. The circularised radius is defined as $r_{\rm circularised}=\sqrt{a b}$, where $a$ and $b$ are the semi-major and semi-minor axis, respectively. The blue shaded region shows the best-fitted core+halo model and the 68 \% confidence interval. For comparison, we also show the surface brightness profiles extracted on the best-fit core and exponential halo components from the two-component modelling in 2D (green and purple lines, respectively). We note that the PSF is similar to the profile of the core-component which is unresolved in the MUSE data. The 1$\sigma$ noise level is shown in grey. }
\label{fig:1D_SB}
\end{figure}

In Fig. $\ref{fig:1D_SB}$ we show the average surface brightness of CR7's Ly$\alpha$ emission as a function of circularised radius. For comparison, we also show the best-fit core and halo models. We then also fit the observed 1D profile as a combination of the core component (an exponential component with effective radius $0.30$ kpc convolved with the Moffat profile of the MUSE PSF; \S $\ref{sec:data}$) and an exponential halo.\footnote{For simplicity, we do not convolve the extended exponential halo-component with the PSF in the one-dimensional fit and we note that this fit agrees very well with the 2D fit that takes the PSF into account (i.e. the purple line in Fig. $\ref{fig:1D_SB}$).} As listed in Table $\ref{tab:lya_fits}$ and shown in Fig. $\ref{fig:1D_SB}$, the best-fit halo scale radius (r$_{\rm s, halo} = 2.9\pm0.2$ kpc) agrees very well with the fitted value using the 2D model.

\section{No neighbouring LAEs in the MUSE data} \label{app:neighbours}
We do not find any neighbouring LAEs with a luminosity $\gtrsim3\times10^{42}$ erg s$^{-1}$ within the FoV of the MUSE data (1 arcmin$^2$; $\approx320$ pkpc$^2$ at $z=6.6$) and between $\lambda_{\rm obs} = 910-930$ nm ($z\approx6.49-6.65$; $\Delta v \approx -4800$ to $+2000$ km s$^{-1}$ with respect to CR7). We estimate the completeness of our line-search using CubEx following the methodology detailed in \cite{Matthee2019MUSE}. In short, we simulate fake LAEs with a truncated-gaussian line profile and a surface brightness profile following the mean profile of LAEs at $z\approx5$ in \cite{Wisotzki2018}. These simulated LAEs are convolved with the LSF and PSF of the data and injected in random positions at $z=6.49-6.65$ in a continuum-subtracted data cube. We then perform recovery experiments with increasing Ly$\alpha$ luminosity and find a 50 \% completeness for a Ly$\alpha$ luminosity of $4 (2.5) \times10^{42}$ erg s$^{-1}$ for an assumed FWHM of 200 (100) km s$^{-1}$.

\bsp	
\label{lastpage}
\end{document}